\documentclass[12pt]{article}
\usepackage{epsfig,amsmath,euscript}

\def\pslash{\rlap{\hspace{0.02cm}/}{p}}
\def\delslash{\rlap{\hspace{0.02cm}/}{\partial}}
\def\lslash{\rlap{/}{l}}

\def\nslash{\rlap{\hspace{0.02cm}/}{n}}
\def\nbslash{\rlap{\hspace{0.02cm}/}{\bar n}}
\def\vslash{\rlap{\hspace{0.02cm}/}{v}}
\def\Dslash{\rlap{\hspace{0.07cm}/}{D}}

\def\calAslash{\rlap{\hspace{0.08cm}/}{{\EuScript A}}}

\def\A{{\EuScript A}}
\def\H{{\EuScript H}}
\def\Q{{\EuScript Q}}
\def\X{{\EuScript X}}

\def\bm#1{\mbox{\boldmath$#1$\unboldmath}}

\begin{document}

\begin{titlepage}

\begin{flushright}
CLNS~03/1849\\
{\tt hep-ph/0311345}\\[0.2cm]
November 26, 2003
\end{flushright}

\vspace{0.7cm}
\begin{center}
\Large\bf 
Factorization and the Soft Overlap Contribution to\\
Heavy-to-Light Form Factors
\end{center}

\vspace{0.8cm}
\begin{center}
{\sc Bj\"orn O.~Lange and Matthias Neubert}\\
\vspace{0.7cm}
{\sl Institute for High-Energy Phenomenology\\
Newman Laboratory for Elementary-Particle Physics, Cornell University\\
Ithaca, NY 14853, U.S.A.}
\end{center}

\vspace{1.0cm}
\begin{abstract}
\vspace{0.2cm}\noindent
Using the formalism of soft-collinear effective theory, a complete 
separation of short- and long-distance contributions to heavy-to-light
transition form factors at large recoil is performed. The universal 
functions $\zeta_M(E)$ parameterizing the ``soft overlap'' contributions 
to the form factors are defined in terms of matrix elements in the 
effective theory. Endpoint configurations corresponding to kinematic 
situations where one of the valence partons in the external mesons 
carries very small momentum are accounted for in terms of operators 
involving soft-collinear messenger fields. They contribute at leading 
order in $\Lambda_{\rm QCD}/E$ and spoil factorization. An analysis of 
operator mixing and renormalization-group evolution in the effective 
theory reveals that the intermediate scale $\sqrt{E\Lambda}$ is without 
significance to the soft functions $\zeta_M(E)$, and that the soft 
overlap contribution does not receive a significant perturbative 
(Sudakov) suppression. 
\end{abstract}
\vfil

\end{titlepage}

\section{Introduction}

$B$-meson form factors describing transitions into light pseudoscalar or 
vector mesons play an essential role in the phenomenology of exclusive 
weak decays. They enter in semileptonic decays such as $B\to\pi\,l\,\nu$, 
which provide access to the CKM matrix element $|V_{ub}|$, and in rare 
decays such as $B\to K^*\gamma$ and $B\to K\,l^+ l^-$, which are 
sensitive probes for physics beyond the Standard Model. Form factors also 
enter in the analysis of exclusive hadronic decays such as $B\to\pi\pi$, 
which give access to the angles $\gamma$ and $\alpha$ of the unitarity 
triangle.

\begin{figure}
\begin{center}
\epsfig{file=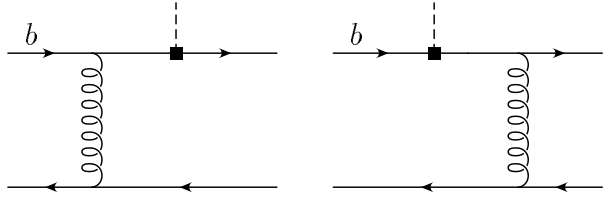,width=8cm} 
\end{center}
\centerline{\parbox{14cm}{\caption{\label{fig:twopart}
Gluon exchange contributions to heavy-to-light form factors. The 
flavor-changing current is denoted by a dashed line. The lines to the 
left belong to the $B$ meson, and those to the right belong to the light 
meson $M$.}}}
\end{figure}

In many of these applications form factors are needed near zero momentum
transfer ($q^2\approx 0$), corresponding to a kinematic situation in 
which a flavor-changing weak current turns a heavy $B$ meson at rest into 
a highly energetic light meson. Such heavy-to-light transitions at high 
recoil are suppressed in QCD by inverse powers of the heavy-quark mass 
$m_b\gg\Lambda_{\rm QCD}$. (This is contrary to the case of 
heavy-to-heavy transitions such as $B\to D$, which are unsuppressed in 
the heavy-quark limit.) The challenge in understanding the physics of 
these processes is to describe properly the transformation of the soft 
constituents of the $B$ meson into the fast moving, collinear 
constituents of the energetic light meson in the final state. At lowest 
order in perturbation theory this transformation can be achieved by the 
exchange of a gluon, as depicted in Figure~\ref{fig:twopart}. Naive power 
counting suggests that the virtualities of the gluon propagators in these 
graphs are parametrically large, of order $E\Lambda$ (with 
$\Lambda\sim\Lambda_{\rm QCD}$), where
\begin{equation}
   E = \frac{m_B^2+m_M^2-q^2}{2m_B}\gg \Lambda
\end{equation}
is the energy of the final-state meson $M$ measured in the $B$-meson 
rest frame. We assume that the partons inside the light meson carry a 
significant fraction of its total energy, and that the light constituents 
of the $B$ meson carry soft momenta of order $\Lambda$. It would then 
appear that the form factor is governed by a hard gluon exchange, which 
can be dealt with using perturbative methods for hard exclusive QCD 
processes \cite{Lepage:1980fj,Efremov:1979qk}. 

However, the situation is indeed far more complicated. In order to 
analyse the graphs in Figure~\ref{fig:twopart}, it is convenient to work 
in the $B$-meson rest frame and choose the outgoing light-meson momentum 
in the $z$-direction. We then define two light-like vectors 
$n^\mu=(1,0,0,1)$ and $\bar n^\mu=(1,0,0,-1)$ satisfying 
$n\cdot\bar n=2$. The meson momentum can be written as 
$p_M^\mu=E n^\mu+O(m_M^2/4E)$, which is nearly light-like. Next, we 
assign incoming momentum $l$ to the light spectator anti-quark inside the 
$B$ meson, and we write the outgoing momenta $p_1$ and $p_2$ of the 
final-state quark and anti-quark that are absorbed by the light meson as
\begin{equation}\label{p1p2}
   p_1^\mu = x_1 E n^\mu + p_\perp^\mu + \dots \,, \qquad
   p_2^\mu = x_2 E n^\mu - p_\perp^\mu + \dots \,,
\end{equation}
where the longitudinal momentum fractions satisfy $x_1+x_2=1$, the 
transverse momenta are $O(\Lambda)$, and the dots represent terms in the 
direction of $\bar n$ that are $O(\Lambda^2/E)$. Assuming that $x_i=O(1)$ 
and keeping only terms of leading order in $\Lambda/E$, a straightforward 
calculation of the two diagrams yields the partonic amplitude
\begin{equation}\label{LO}
   A = -g^2 \left( \Gamma\,
   \frac{m_b(1+\vslash) - x_2 E\nslash}
        {4m_b E^2 x_2^2\,n\cdot l}\,\gamma_\mu
   + \gamma_\mu\,\frac{E\nslash-\lslash}{4E^2 x_2(n\cdot l)^2}\,
   \Gamma \right) t_a * \gamma^\mu t_a + \dots \,,
\end{equation}
where $v^\mu=(1,0,0,0)$ is the $B$-meson velocity, $\Gamma$ denotes the 
Dirac structure of the flavor-changing current, the $*$ product means 
that the two factors must be sandwiched between quark spinors, and the 
dots represent power-suppressed terms. After projection onto the external 
meson states, the partonic amplitude must be convoluted with light-cone 
distribution amplitudes (LCDAs) governing the distributions of $x_2$ and 
$n\cdot l$ inside the meson states. For $x_2=O(1)$, the terms in the 
parenthesis scale like $1/(E^2\Lambda)$, which after projection onto the 
meson states implies a power suppression of the corresponding 
heavy-to-light form factors. (The contribution in the second term that 
superficially scales like $1/(E\Lambda^2)$ vanishes by the equations of 
motion.) However, the singularities of the amplitude arising for 
$x_2\to 0$ or $n\cdot l\to 0$ are such that the convolution integrals 
with the LCDAs diverge at the endpoints 
\cite{Szczepaniak:1990dt,Burdman:hg}. Near the endpoints the perturbative 
analysis of the diagrams in Figure~\ref{fig:twopart} breaks down, since 
the gluon propagators are no longer far off-shell. This observation is 
the basis for the notion of a soft overlap contribution to the form 
factor (Feynman mechanism), which would formally be the leading 
contribution, since it is not suppressed by a perturbative coupling 
constant $\alpha_s$. The precise nature of this contribution and its 
scaling properties in the heavy-quark limit remain however unclear. 

The QCD factorization approach to exclusive hadronic $B$ decays assumes
that  heavy-to-light form factors are dominated by the soft overlap 
contribution and are truly non-perturbative objects, which are treated 
as hadronic input parameters \cite{Beneke:1999br}. However, it has 
sometimes been argued that the summation of large Sudakov logarithms 
associated with the soft gluon exchange mechanism may lead to a strong 
suppression of the soft overlap term, essentially reinstating the 
perturbative nature of the form factors (see, e.g., 
\cite{Akhoury:uw,Dahm:1995ne,Kurimoto:2001zj}). The assumption of a 
short-distance nature of heavy-to-light form factors at large recoil is 
the basis of the pQCD approach to exclusive hadronic $B$ decays 
\cite{Keum:2000ph}, which is often considered a competitor to QCD 
factorization. Leaving aside some conceptual problems associated with 
this treatment \cite{Descotes-Genon:2001hm}, the issue of Sudakov 
logarithms is an intricate one, because contributions to the form factors 
can arise from several different energy scales. Besides the hard scale 
$E\sim m_B$ and the hadronic scale $\Lambda$, interactions between soft 
and collinear partons involve an intermediate ``hard-collinear'' scale of 
order $\sqrt{E\Lambda}$. To settle the question of Sudakov suppression of 
the soft overlap contribution is one of the main motivations for our 
work.

Recently, much progress in the understanding of exclusive $B$ decays has
been achieved by combining methods developed for hard exclusive processes
with concepts of heavy-quark expansions and effective field theory. 
Soft-collinear effective theory (SCET) has been established as a tool to 
systematically disentangle the various momentum regions and scales 
relevant in processes involving soft and collinear partons 
\cite{Bauer:2000ew,Bauer:2000yr,Bauer:2001yt,Chay:2002vy,Beneke:2002ph,%
Hill:2002vw}. This approach has already been applied to study some 
aspects of heavy-to-light form factors at large recoil 
\cite{Beneke:2002ph,Hill:2002vw,Bauer:2002aj,Beneke:2003xr,Lange:2003jz}. 
Because form factors are power-suppressed quantities, the analysis in the 
framework of SCET is highly non-trivial. Our main goal in the present 
work is to provide a rigorous definition of the soft overlap contribution 
in terms of SCET operators. This will allow us to study systematically 
its dependence on the large scale $E$.

Let us start with a summary of related work by previous authors. The 
discussion of heavy-to-light form factors simplifies in the heavy-quark 
limit $E\gg\Lambda$ and can be summarized by the factorization formula 
\cite{Beneke:2000wa}
\begin{equation}\label{ff}
   f_i^{B\to M}(E) = C_i(E,\mu_{\rm I})\,\zeta_M(\mu_{\rm I},E)
   + T_i(E,\sqrt{E\Lambda},\mu)\otimes\phi_B(\mu)\otimes\phi_M(\mu)
   + \dots ,
\end{equation}
where the dots represent terms that are of subleading order in 
$\Lambda/E$. $C_i$ and $T_i$ are calculable short-distance coefficient 
functions, $\phi_B$ and $\phi_M$ denote the leading-order LCDAs of the 
$B$ meson and the light meson $M$, and the $\otimes$ products imply 
convolutions over light-cone momentum fractions. The functions $\zeta_M$ 
denote universal form factors that only depend on the nature of the light 
final-state meson but not on the Lorentz structure of the currents whose 
matrix elements define the various form factors. The first term in the 
factorization formula therefore implies spin-symmetry relations between 
different form factors, which were first derived in \cite{Charles:1998dr} 
by considering the large-energy limit of QCD. The second term in 
(\ref{ff}), which arises from hard-collinear gluon exchange, breaks these 
symmetry relations \cite{Beneke:2000wa}. 

The arguments $E$ and $\sqrt{E\Lambda}$ in the short-distance 
coefficients $C_i$ and $T_i$ are representative for any of the hard or 
hard-collinear scales in the problem, respectively. The form of the 
factorization formula shown in (\ref{ff}) assumes that the factorization 
scale $\mu_{\rm I}$ in the first term is chosen to lie between the hard 
scale $E$ and the hard-collinear scale $\sqrt{E\Lambda}$, whereas the 
scale $\mu$ in the second term is chosen to lie below the hard-collinear 
scale. The Wilson coefficients $C_i$ receive contributions at tree level, 
whereas the hard-scattering kernels $T_i$ start at first order in 
$\alpha_s(\sqrt{E\Lambda})$. Naively, one would conclude that the 
spin-symmetry preserving term provides the leading contribution to the 
form factors in the heavy-quark limit. Only then the notion of an 
approximate spin symmetry would be justified. However, the situation is 
more complicated because as written in (\ref{ff}) the universal functions 
$\zeta_M(\mu_{\rm I},E)$ still depend on the short-distance scale 
$\sqrt{E\Lambda}$; in fact, their $E$-dependence remains unspecified. 
Contrary to the hard-scattering term, for which all short-distance 
physics has been factorized into the kernels $T_i$, the first term in 
(\ref{ff}) has so far not been factorized completely. This is reflected 
in existing discussions of the factorization properties of heavy-to-light 
form factors in the context of SCET 
\cite{Beneke:2002ph,Hill:2002vw,Bauer:2002aj,Beneke:2003xr}. The matching 
of a QCD form factor onto matrix elements in SCET is usually done in two 
steps: QCD\,$\to$\,SCET$_{\rm I}$\,$\to$\,SCET$_{\rm II}$. In the first 
step, hard fluctuations with virtualities on the scale $E\sim m_B$ are 
integrated out, while in the second step hard-collinear modes with 
virtualities of order $\sqrt{E\Lambda}$ are removed. Both steps of this 
matching are understood for the hard-scattering term in the factorization 
formula, for which the kernels $T_i$ can be factorized as 
\cite{Bauer:2002aj}
\begin{equation}\label{Tifact}
   T_i(E,\sqrt{E\Lambda},\mu) = \sum_j H_{ij}(E,\mu)
   \otimes J_j(\sqrt{E\Lambda},\mu) \,. \vspace{-0.3cm}
\end{equation}
On the other hand, the matching onto SCET$_{\rm II}$ has not yet been 
performed for the universal functions $\zeta_M(E,\mu_{\rm I})$ entering 
the first term.\footnote{The notation $\mu_{\rm I}$ serves as a reminder 
that the factorization scale for this term is defined in the intermediate 
effective theory SCET$_{\rm I}$.}  
In \cite{Bauer:2002aj,Beneke:2003xr} these functions are defined in terms 
of matrix elements in the intermediate effective theory SCET$_{\rm I}$, 
which leaves open the possibility that they could be dominated by 
short-distance physics. In the present work we show that the functions 
$\zeta_M$ renormalized at a hard-collinear scale 
$\mu_{hc}\sim\sqrt{E\Lambda}$ can be factorized further according to
\begin{equation}\label{scet2ff}
   \zeta_M(\mu_{hc},E) = \sum_k D_k^{(M)}(\sqrt{E\Lambda},\mu_{hc},\mu)
   \otimes\xi_k^{(M)}(\mu,E) \,, \vspace{-0.3cm}
\end{equation}
where the functions $\xi_k^{(M)}$ are defined in terms of hadronic matrix 
elements of SCET$_{\rm II}$ operators. By solving the 
renormalization-group (RG) equation for these operators we show that the 
universal form factors $\zeta_M$ do not receive a significant 
perturbative suppression, neither by a power of 
$\alpha_s(\sqrt{E\Lambda})$ nor by resummed Sudakov logarithms. This 
statement holds true even in the limit where $m_B$ is taken to be much 
larger than the physical $B$-meson mass. Perhaps somewhat surprisingly, 
we find that the hard-collinear scale $\sqrt{E\Lambda}$ is without 
physical significance to the soft functions $\zeta_M$. Switching from 
SCET$_{\rm I}$ to SCET$_{\rm II}$ one describes the same physics using a 
different set of degrees of freedom. This observation leads us to the 
most important conclusion of our paper, namely that the long-distance, 
soft overlap contribution to heavy-to-light form factors exists. However, 
we point out that even at low hadronic scales $\mu\sim\Lambda$ the 
functions $\xi_k^{(M)}$ contain a dependence on the large recoil energy 
$E$ which is of long-distance nature and cannot be factorized using RG 
techniques. As a result, it is impossible to determine the asymptotic 
behavior of the QCD form factors $f_i^{B\to M}$ as $E\to\infty$ using 
short-distance methods.

Let us comment in more detail on the philosophy adopted by Bauer et al.\ 
in \cite{Bauer:2002aj}, where an attempt was made to prove the 
factorization formula (\ref{ff}) based on hard-collinear/soft 
factorization of operators in SCET$_{\rm I}$. In this interesting paper, 
the authors introduced the notion of ``factorizable operators'', in which 
all soft gluon interactions occur in redefined soft quark fields 
${\cal H}=Y^\dagger h$ and ${\cal Q}=Y^\dagger q$, which in soft 
light-cone gauge coincide with the original fields for heavy and light 
quarks. (Here $Y$ is a soft Wilson line in SCET$_{\rm I}$.) They argued 
that the $B\to M$ matrix elements of ``factorizable'' SCET$_{\rm I}$ 
operators could be written in terms of convergent convolution integrals 
of hard-scattering kernels with leading-order LCDAs, and that endpoint 
divergences of convolution integrals only arise in an attempt to evaluate 
the matrix elements of ``non-factorizable'' operators. They also showed 
that the operators which give rise to spin-symmetry breaking 
contributions to the form factors are all ``factorizable''. The matrix 
elements of the ``non-factorizable'' SCET$_{\rm I}$ operators were used 
to define the universal functions $\zeta_M$. In the present paper we show 
that hard-collinear/soft factorization in SCET$_{\rm I}$ alone is not 
sufficient to guarantee the absence of endpoint singularities in the 
convolution integrals that enter the factorization formula. By completing 
the matching of SCET$_{\rm I}$ operators onto the final effective theory 
SCET$_{\rm II}$, we identify a particular ``factorizable'' SCET$_{\rm I}$ 
operator (called $T_0^{\rm F}$ in \cite{Bauer:2002aj}) whose matrix 
elements cannot be written in the form of convergent convolution 
integrals of perturbative kernels with LCDAs. 

To understand the physics of endpoint singularities it is necessary to 
perform the matching onto SCET$_{\rm II}$ explicitly and study the 
interplay of the various fields that are contained in this effective 
theory. For some time it was believed that SCET$_{\rm II}$ factorizes 
into disconnected soft and collinear sectors. If true, this would imply 
the absence of endpoint singularities, because they correspond to 
momentum scalings that are not accounted for by soft or collinear fields. 
It would also imply that the functions $\zeta_M$ are proportional to a 
power of $\alpha_s(\sqrt{E\Lambda})$ (modulo logarithms), since a 
hard-collinear gluon exchange is required to transform the soft spectator 
anti-quark inside the $B$ meson into a collinear anti-quark that can be 
absorbed by the final-state light meson $M$. As a result, one would 
expect that the spin-symmetry relations could receive $O(1)$ corrections 
that are not suppressed by any small parameter. Recently, however, we 
have argued that soft/collinear factorization is {\em not\/} a generic 
property of SCET$_{\rm II}$, not even at leading order in power counting 
\cite{Becher:2003qh,Becher:2003kh}. A non-perturbative cross-talk between 
the two sectors of the theory exists, which can be dealt with most 
elegantly by introducing a new kind of fields called soft-collinear 
messengers.\footnote{This cross-talk is also seen in the approach of 
\cite{Beneke:2003xr}, where it appears as a dependence of soft and 
collinear contributions on a common analytic regulator introduced to 
define certain one-loop graphs with light-like external momenta that 
cannot be regularized dimensionally.}
In the present work we use this framework and apply it to the case of 
heavy-to-light form factors. We show that the endpoint singularities 
arising in convolution integrals are removed by including the 
contributions from soft-collinear fields. The sum of all SCET$_{\rm II}$
matrix elements is free of such singularities. Even at hadronic scales
$\mu\sim\Lambda$, the functions $\zeta_M$ cannot be written as 
convolution integrals with LCDAs, meaning that soft and collinear 
contributions cannot be factorized. 

In order to prove the factorization formula (\ref{ff}) one needs to show
that all contributions to the form factors that do not obey spin-symmetry 
relations can be written in terms of convolution integrals involving the 
leading-order LCDAs of the $B$ meson and the light meson. Specifically, 
this means showing that (i) no higher Fock states (or higher-twist 
two-particle distribution amplitudes) contribute at leading power, and 
(ii) the convolution integrals are convergent to all orders in 
perturbation theory. Point (i) can be dealt with using the power-counting 
rules of SCET along with reparameterization invariance 
\cite{Bauer:2002aj,Beneke:2003xr}. Here we use our formalism to complete 
step (ii) of the factorization proof. (The convergence of the convolution 
integral over the $B$-meson LCDA can also be shown using arguments along 
the lines of \cite{Bosch:2003fc,Lange:2003ff}.)

The remainder of the paper is structured as follows: After a brief
introduction to the SCET$_{\rm II}$ formalism in Section~\ref{sec:scet}, 
we discuss in Section~\ref{sec:matching} the matching of the universal
functions $\zeta_M$ onto operators in the low-energy effective theory 
containing only soft and collinear fields. We construct a basis of the 
operators contributing at leading power and derive the tree-level 
expressions for their Wilson coefficient functions. We find that the 
``factorizable'' SCET$_{\rm I}$ operator $T_0^{\rm F}$ (in the language 
of \cite{Bauer:2002aj}) matches onto some SCET$_{\rm II}$ operators whose 
matrix elements cannot be written as convergent convolutions over 
$B$-meson and light-meson LCDAs. These matrix elements are associated 
with non-valence Fock states of the external mesons, which contribute at 
the same order in power counting as the two-particle valence Fock states 
\cite{Hill:2002vw,Lange:2003jz}. The appearance of endpoint singularities 
in convolution integrals signals the relevance of contributions involving 
long-distance soft-collinear messenger modes. This connection is made 
more explicit in Section~\ref{sec:toy}, where we analyze in detail a toy 
model of endpoint singularities. In Section~\ref{sec:softoverlap} we then 
complete the parameterization of the universal functions $\zeta_M$ in 
terms of SCET$_{\rm II}$ matrix elements by including the messenger 
contributions. The RG mixing of the various SCET$_{\rm II}$ operators is 
discussed in Section~\ref{sec:Sudakov}. We show that the linear 
combination of operators relevant to the form factors is an eigenvector 
of the RG evolution matrix and identify the corresponding eigenvalue with 
the anomalous dimension of a current operator. Solving the evolution 
equation for this operator, we resum the short-distance Sudakov 
logarithms arising in the evolution from the hard scale $\mu\sim E$ down 
to hadronic scales $\mu\sim\Lambda$ at next-to-leading order. Finally, in 
Section~\ref{sec:fact} we apply our formalism to prove that the 
spin-symmetry violating term in the factorization formula (\ref{ff}) is 
free of endpoint singularities. Our results are summarized in 
Section~\ref{sec:concl}.

\section{Soft-collinear effective theory}
\label{sec:scet}

The construction of the effective Lagrangian for SCET$_{\rm II}$ has been 
discussed in detail in \cite{Hill:2002vw,Becher:2003qh,Becher:2003kh}, to 
which we refer the reader for details. Here we summarize the main results 
and notations needed for our discussion.

In order to account for the fact that different components of particle 
momenta and fields scale differently with the large scale $E$, one 
decomposes 4-vectors in the light-cone basis constructed with the help of 
the two light-like vectors $n^\mu$ and $\bar n^\mu$. Specifically, 
$n^\mu=(1,0,0,1)$ is chosen to be the direction of an outgoing fast 
hadron (or a jet of hadrons), and $\bar n^\mu=(1,0,0,-1)$ points in the 
opposite direction. The light-cone decomposition of a 4-vector reads
\begin{equation}
   p^\mu = (n\cdot p)\,\frac{\bar n^\mu}{2}
    + (\bar n\cdot p)\,\frac{n^\mu}{2} + p_\perp^\mu
   \equiv p_+^\mu + p_-^\mu + p_\perp^\mu \,,
\end{equation}
where $p_\perp\cdot n=p_\perp\cdot\bar n=0$. The relevant SCET$_{\rm II}$ 
degrees of freedom describing the partons in the external hadronic states 
of exclusive $B$ decays are soft or collinear. Introducing an expansion
parameter $\lambda=\Lambda/E\ll 1$, we have 
$p_s^\mu\sim E(\lambda,\lambda,\lambda)$ for soft momenta and 
$p_c^\mu\sim E(\lambda^2,1,\lambda)$ for collinear momenta, where we 
indicate the scaling properties of the components 
$(n\cdot p,\bar n\cdot p,p_\perp)$. The corresponding effective-theory 
fields and their scaling properties are $h\sim\lambda^{3/2}$ (soft heavy 
quark), $q_s\sim\lambda^{3/2}$ (soft light quark), 
$A_s^\mu\sim(\lambda,\lambda,\lambda)$ (soft gluon), and 
$\xi\sim\lambda$ (collinear quark), $A_c^\mu\sim(\lambda^2,1,\lambda)$ 
(collinear gluon). The effective theory also contains soft-collinear 
quark and gluon messenger fields, $\theta\sim\lambda^2$ and 
$A_{sc}^\mu\sim(\lambda^2,\lambda,\lambda^{3/2})$, which have couplings 
to both soft and collinear fields. 

The various fields present in the effective theory obey a set of 
``homogeneous'' gauge transformations, which leave their scaling 
properties unchanged \cite{Bauer:2001yt,Becher:2003qh,Beneke:2002ni}. In 
these transformations the soft-collinear gluon field is treated as a 
background field. Soft fields transform under soft and soft-collinear 
gauge transformations but are invariant under collinear transformations. 
Likewise, collinear fields transform under collinear and soft-collinear 
gauge transformations but are invariant under soft transformations. 
Soft-collinear fields only transform under soft-collinear gauge 
transformations.

The effective Lagrangian of SCET$_{\rm II}$ can be split up as
\begin{equation}\label{Lscet}
   {\cal L}_{\rm SCET} = {\cal L}_s + {\cal L}_c + {\cal L}_{sc}
   + {\cal L}_{\rm int}^{(0)} + \dots \,,
\end{equation}
where the dots represent power-suppressed interactions. None of the terms 
in the Lagrangian is renormalized beyond the usual renormalization of the 
fields and coupling constant. The integration measure $d^4x$ in the 
action $S_{\rm SCET}=\int d^4x\,{\cal L}_{\rm SCET}$ scales like 
$\lambda^{-4}$ for all terms except the soft-collinear Lagrangian 
${\cal L}_{sc}$, for which it scales like $\lambda^{-6}$. The first three 
terms above correspond to the Lagrangians of soft particles (including 
heavy quarks), collinear particles, and soft-collinear particles. They 
are given by
\begin{equation}
\begin{aligned}
   {\cal L}_s &= \bar q_s\,i\Dslash_s\,q_s + \bar h\,iv\cdot D_s\,h
    + {\cal L}_s^{\rm glue} \,, \\[0.1cm]
   {\cal L}_c
   &= \bar\xi\,\frac{\nbslash}{2}\,in\cdot D_c\,\xi
    - \bar\xi\,i\Dslash_{c\perp}\,\frac{\nbslash}{2}\,
    \frac{1}{i\bar n\cdot D_c}\,i\Dslash_{c\perp}\,\xi 
    + {\cal L}_c^{\rm glue} \,, \\[-0.1cm]
   {\cal L}_{sc}
   &= \bar\theta\,\frac{\nbslash}{2}\,in\cdot D_{sc}\,\theta
    - \bar\theta\,i\Dslash_{sc\perp}\,\frac{\nbslash}{2}\,
    \frac{1}{i\bar n\cdot D_{sc}}\,i\Dslash_{sc\perp}\,\theta
    + {\cal L}_{sc}^{\rm glue} \,,
\end{aligned}
\end{equation}
where $iD_s^\mu\equiv i\partial^\mu+g A_s^\mu$ is the covariant 
derivative built using soft gauge fields, etc. Collinear, soft-collinear, 
and heavy-quark fields in the effective theory are described by 
two-component spinors subject to the constraints $\nslash\,\xi=0$, 
$\nslash\,\theta=0$, and $\vslash\,h=h$. The gluon Lagrangians in the 
three sectors retain the same form as in full QCD,  but with the gluon 
fields restricted to the corresponding subspaces of their soft, 
collinear, or soft-collinear Fourier modes. 

In interactions with other fields, the soft-collinear fields (but not the 
soft and collinear fields) are multipole expanded about the points $x_-$
(interactions with collinear fields) or $x_+$ (interactions with soft 
fields). The leading-order interactions are given by
\begin{eqnarray}\label{Lint0}
   {\cal L}_{\rm int}^{(0)}(x)
   &=& \bar q_s(x)\,\frac{\nslash}{2}\,g\bar n\cdot A_{sc}(x_+)\,q_s(x)
    + \bar h(x)\,\frac{n\cdot v}{2}\,g\bar n\cdot A_{sc}(x_+)\,h(x)
    \nonumber\\
   &+& \bar\xi(x)\,\frac{\nbslash}{2}\,g n\cdot A_{sc}(x_-)\,
    \xi(x) + \mbox{pure glue terms} \,.
\end{eqnarray}
Momentum conservation implies that soft-collinear fields can only couple 
to either soft or collinear fields, but not both. The gluon 
self-couplings will not be relevant to our discussion. 

At leading order in $\lambda$, matching calculations can be done by 
evaluating diagrams with external quark and gluon states and making the 
replacements
\begin{equation}
\begin{aligned}
   \psi_\xi &\to W_c^\dagger\,\xi \,, \qquad
    \psi_q\to S_s^\dagger\,q_s \,, \qquad
    \psi_Q\to S_s^\dagger\,h \,, \\
   g A_{c\perp}^\mu &\to W_c^\dagger\,(iD_{c\perp}^\mu W_c) \,, \qquad
    g A_{s\perp}^\mu\to S_s^\dagger\,(iD_{s\perp}^\mu S_s) \,,
\end{aligned}
\end{equation}
where
\begin{equation}\label{WSdef}
\begin{aligned}
   W_c(x) &= \mbox{P}\exp\left( ig\int_{-\infty}^0 dt\,
    \bar n\cdot A_c(x+t\bar n) \right) , \\
   S_s(x) &= \mbox{P}\exp\left( ig\int_{-\infty}^0 dt\,
    n\cdot A_s(x+tn) \right)
\end{aligned} 
\end{equation}
are collinear and soft Wilson lines \cite{Bauer:2001yt,Beneke:2002ph}, 
which effectively put the SCET fields into light-cone gauge. The 
leading-order interactions between soft-collinear fields and soft or 
collinear fields in (\ref{Lint0}) can be removed by a redefinition of the 
soft and collinear fields, which is analogous to the decoupling of soft 
gluons in SCET$_{\rm I}$ \cite{Bauer:2001yt}. To this end, one 
rewrites the fields arising in matching calculations in terms of the
gauge-invariant building blocks
\begin{equation}\label{blocks}
\begin{aligned}
   W_c^\dagger\,\xi &= S_{sc}(x_-)\,\X \,, \qquad
    S_s^\dagger\,q_s = W_{sc}(x_+)\,\Q_s \,, \qquad
    S_s^\dagger\,h = W_{sc}(x_+)\,\H \,, \\
   W_c^\dagger\,(iD_{c\perp}^\mu W_c)
   &= S_{sc}(x_-)\,\A_{c\perp}^\mu\,S_{sc}^\dagger(x_-) \,, \qquad
    S_s^\dagger\,(iD_{s\perp}^\mu S_s)
    = W_{sc}(x_+)\,\A_{s\perp}^\mu\,W_{sc}^\dagger(x_+) \,,
\end{aligned}
\end{equation}
where fields without argument live at position $x$. The quantities 
$W_{sc}$ and $S_{sc}$ are a new set of Wilson lines defined in analogy 
with $W_c$ and $S_s$ in (\ref{WSdef}), but with the gluon fields replaced 
by soft-collinear gluon fields in both cases. The calligraphic fields in
(\ref{blocks}) are invariant under soft, collinear, and soft-collinear 
gauge transformations. When they are introduced in the SCET$_{\rm II}$ 
Lagrangian the leading-order interaction terms denoted by 
${\cal L}_{\rm int}^{(0)}$ vanish. Residual interactions between 
soft-collinear fields and soft or collinear fields start at 
$O(\lambda^{1/2})$ \cite{Becher:2003qh}. We will study some of these 
interactions in more detail in Section~\ref{sec:softoverlap}.

\section{Matching calculations}
\label{sec:matching}

In the matching of the intermediate effective theory SCET$_{\rm I}$ onto 
the final low-energy effective theory SCET$_{\rm II}$, hard-collinear 
modes with virtuality of order $\sqrt{E\Lambda}$ are integrated out, and 
their effects are included in short-distance coefficient functions. Our 
primary goal in this section is to construct a basis of operators 
relevant to the matching of the universal functions $\zeta_M$ in 
(\ref{ff}) onto the low-energy theory (at leading power in $\Lambda/E$), 
and to calculate their Wilson coefficients at lowest order in 
perturbation theory. As always, matching calculations can be done using 
on-shell external quark and gluon states rather than the physical meson 
states, whose matrix elements define the form factors. The results for 
the Wilson coefficients are insensitive to infra-red physics. At the end 
of this section we briefly discuss also the spin-symmetry breaking 
contributions to the form factors. 

\subsection{Spin-symmetric contributions}

Heavy-to-light form factors are defined in terms of $B\to M$ matrix 
elements of flavor-changing currents $\bar q\,\Gamma\,b$. The 
spin-symmetric contributions to the form factors are characterized by the 
fact that in the intermediate effective theory (i.e., after hard 
fluctuations with virtuality $\mu\sim E\sim m_B$ are integrated out) they 
contain the Dirac structure $\Gamma$ sandwiched between the two projection
operators $\frac14\nbslash\nslash$ and $\frac12(1+\vslash)$. This implies 
that $\Gamma$ can be decomposed into a linear combination of only three 
independent Dirac structures, which leads to symmetry relations between 
various form factors \cite{Charles:1998dr}. It has been shown in 
\cite{Bauer:2002aj} that in SCET$_{\rm I}$ the relevant operators can be 
written as time-ordered products of the effective Lagrangian with 
effective current operators 
$J_M^{(0)}(x)=(\bar\xi_{hc} W_{hc})(x)\,\Gamma_M\,h(x_-)$ defined by the 
matching relation
\begin{equation}\label{current}
   \bar q\,\Gamma\,b = \sum_M K_M^\Gamma(m_b,E,\mu)\,J_M^{(0)}
   + \dots \,, \vspace{-0.3cm}
\end{equation}
where the dots denote power-suppressed terms. Here 
$\Gamma_M=1,\gamma_5,\gamma_{\perp\nu}$ is one of the three Dirac basis 
matrices that remain after the projections onto the two-component spinors
$\bar\xi_{hc}$ and $h$. The variable $E$ entering the coefficients 
$K_M^\Gamma$ is the total energy carried by collinear particles (more 
precisely, $2E=\bar n\cdot p_c^{\rm tot}$), which in our case coincides 
with the energy of the final-state meson $M$. Since the Lagrangian is a 
Lorentz scalar, it follows that for $B\to M$ transitions each of the 
three possibilities corresponds to a particular choice of the final-state 
meson (hence the label ``$M$'' on $\Gamma_M$), namely $\Gamma_M=1$ for 
$M=P$ a pseudoscalar meson, $\Gamma_M=\gamma_5$ for $M=V_\parallel$ a
longitudinally polarized vector meson, and $\Gamma_M=\gamma_{\perp\nu}$
for $M=V_\perp$ a transversely polarized vector meson. The coefficients
$K_M^\Gamma$ for the various currents that are relevant in the discussion 
of heavy-to-light form factors are summarized in Table~\ref{tab:coefs}, 
where we use the definitions
\begin{equation}
   g_\perp^{\mu\nu} = g^{\mu\nu}
    - \frac{n^\mu\bar n^\nu+n^\nu\bar n^\mu}{2} \,, \qquad
   \epsilon_\perp^{\mu\nu} = \frac12\,\epsilon^{\mu\nu\alpha\beta} 
    \bar n_\alpha n_\beta
\end{equation}
for the symmetric and anti-symmetric tensors in the transverse plane. (We 
use $\epsilon_{0123}=1$ and $\gamma_5=i\gamma^0\gamma^1\gamma^2\gamma^3$.)
Once a definition for the heavy-to-light form factors is adopted, it is 
an easy exercise to read off from the table which of the coefficient 
functions $C_i$ contribute to a given form factor. This determines the 
functions $C_i(E,\mu_{\rm I})$ in (\ref{ff}). (In general, these 
functions are linear combinations of the $C_i$ in Table~\ref{tab:coefs}.)

\begin{table}
\centerline{\parbox{14cm}{\caption{\label{tab:coefs}
Coefficients $K_M^\Gamma$ arising in the leading-order matching of 
flavor-changing currents from QCD onto SCET$_{\rm I}$. The coefficients 
$C_i$ are defined in \cite{Bauer:2000yr}. We denote $\hat q=q/m_B$ and 
$\hat E=E/m_B$, where $q=p_B-p_M$.}}}
\vspace{0.1cm}
\begin{center}
\begin{tabular}{|c|ccc|}
\hline\hline
Current & $M=P$ & $M=V_\parallel$ & $M=V_\perp$ \\[-0.1cm]
$\bar q\,\Gamma\,b$ & ($\Gamma_M=1$) & ($\Gamma_M=\gamma_5$)
 & ($\Gamma_M=\gamma_{\perp\nu}$) \\
\hline\hline
$\bar q\gamma^\mu b$ & $(n^\mu C_4 + v^\mu C_5)$ & ---
 & $g_\perp^{\mu\nu} C_3$ \\
$\bar q\gamma^\mu\gamma_5 b$ & --- & $-(n^\mu C_7 + v^\mu C_8)$
 & $-i\epsilon_\perp^{\mu\nu} C_6$ \\
$\bar q\,i\sigma^{\mu\nu}\hat q_\nu b$
 & $[v^\mu - (1-\hat E) n^\mu]\,C_{11}$ & ---
 & $-g_\perp^{\mu\nu} [C_9 + (1-\hat E) C_{12}]$ \\
$\bar q\,i\sigma^{\mu\nu}\gamma_5\hat q_\nu b$
 & --- & $[v^\mu - (1-\hat E) n^\mu]\,C_{10}$
 & $-i\epsilon_\perp^{\mu\nu} (C_9 + \hat E C_{12})$ \\
\hline\hline
\end{tabular}
\end{center}
\vspace{-0.2cm}
\end{table}

Our goal is to match the time-ordered product 
$i\int d^4x\,\mbox{T}\,\{ J_M^{(0)}(0),{\cal L}_{\rm SCET_{\rm I}}(x)\}$ 
onto operators in SCET$_{\rm II}$, focusing first on operators which 
include soft and collinear fields only. The insertion of a subleading 
interaction from the SCET$_{\rm I}$ Lagrangian is required to transform 
the soft $B$-meson spectator anti-quark into a hard-collinear final-state 
parton. In the resulting SCET$_{\rm II}$ interaction terms the soft and 
collinear fields must be multipole expanded, since their momenta have 
different scaling properties with the expansion parameter $\lambda$ 
\cite{Beneke:2002ph}. Soft fields are expanded about $x_+=0$ while 
collinear ones are expanded about $x_-=0$. In general, collinear fields 
can live on different $x_+$ positions while soft fields can live on 
different $x_-$ positions. The resulting non-localities of 
SCET$_{\rm II}$ operators are allowed by gauge invariance and are needed 
to reproduce the dependence of decay amplitudes on the longitudinal 
momentum components $\bar n\cdot p_c$ and $n\cdot p_s$ of collinear and 
soft momenta. The relevant operators can be written as matrix elements of 
color singlet-singlet four-quark operators multiplied by 
position-dependent Wilson coefficients, i.e. 
\cite{Hill:2002vw,Becher:2003kh}
\begin{eqnarray}
   \int ds\,dt\,\widetilde D_k(s,t,E,\mu)
   &&\!\!\! [(\bar\xi\,W_c)(x_+ +x_\perp)\dots
    (W_c^\dagger\,\xi)(x_+ +x_\perp+s\bar n)] \nonumber\\[-0.2cm]
   \times &&\!\!\! [(\bar q_s\,S_s)(x_- +x_\perp+tn)\dots
    (S_s^\dagger\,h)(x_- +x_\perp)] \,,
\end{eqnarray}
where the dots represent different Dirac structures. There is no need to 
include color octet-octet operators, since they have vanishing 
projections onto physical hadron states and do not mix into the color 
singlet-singlet operators under renormalization. The Fourier transforms 
of the coefficient functions defined as
\begin{equation}
   D_k(\omega,u_2,E,\mu) = \int ds\,e^{2iE u_2 s}
   \int dt\,e^{-i\omega t}\,\widetilde D_k(s,t,E,\mu) 
\end{equation}
coincide with the momentum-space Wilson coefficient functions, which we
will calculate below. (In the matching calculation, $\omega$ is 
identified with the component $n\cdot l$ of the incoming spectator 
momentum, and $u_2$ is identified with the longitudinal momentum fraction 
$x_2$ carried by the collinear anti-quark in the final-state meson.) 
Consider now what happens when we rewrite the operators above in terms of 
the gauge-invariant building blocks defined in (\ref{blocks}). Since the 
collinear fields are evaluated at $x_-=0$ and the soft fields at $x_+=0$, 
it follows that all factors of the soft-collinear Wilson lines 
$S_{sc}(0)$ and $W_{sc}(0)$ cancel out. Hence, we can rewrite the 
operators in the form (setting $x=0$ for simplicity)
\begin{equation}
   \int ds\,dt\,\widetilde D_k(s,t,E,\mu)\,
   [\bar\X(0)\dots\X(s\bar n)]\,[\bar\Q_s(tn)\dots\H(0)] \,.
\end{equation}

At leading order in $\lambda$, operators can also contain insertions of 
transverse derivatives or gauge fields between the collinear or soft 
quark fields. Since $\partial_\perp^\mu\sim\lambda$ and 
$\A_{c\perp}^\mu\sim\A_{s\perp}^\mu\sim\lambda$, such transverse 
insertions must be accompanied by a factor of 
$\nslash/in\cdot\partial_s\sim\lambda^{-1}$. The inverse derivative 
operator acts on a light soft field and implies an integration over the 
position of that field on the $n$ light-cone, the effect of which can be 
absorbed into the Wilson coefficient functions. The appearance of 
$\nslash$ in the numerator is enforced by reparameterization invariance. 
Using arguments along the lines of \cite{Hill:2002vw,Bosch:2003fc} it 
then follows that in our case only single insertions of transverse 
objects are allowed. Finally, the multipole expansion of the soft fields 
in SCET$_{\rm I}$ ensures that the component $\bar n\cdot p_s$ of soft 
momenta does not enter Feynman diagrams at leading power. Likewise, at 
leading power there are no operators that contain 
$in\cdot\partial_c\sim\lambda^2$ (acting on collinear fields) or 
$n\cdot\A_c\sim\lambda^2$, since these are always suppressed with respect 
to the corresponding transverse quantities. In the presence of additional 
transverse derivatives or gluon fields the above argument about the 
cancellation of soft-collinear Wilson lines under the field redefinition 
(\ref{blocks}) remains true, so that the resulting operators can again be 
expressed in terms of gauge-invariant building blocks. The new structures 
containing gluon fields are of the form
\begin{equation}
\begin{aligned}
   &\int dr\,ds\,dt\,\widetilde D_k(r,s,t,E,\mu)\,
    [\bar\X(0)\dots\A_{c\perp}^\mu(r\bar n)\dots\X(s\bar n)]\,
    [\bar\Q_s(tn)\dots\H(0)] \,, \\
   &\int ds\,dt\,du\,\widetilde D_k(s,t,u,E,\mu)\,
    [\bar\X(0)\dots\X(s\bar n)]\,
    [\bar\Q_s(tn)\dots\A_{s\perp}^\mu(un)\dots\H(0)] \,,
\end{aligned}
\end{equation}
and we define the corresponding Fourier-transformed coefficient functions 
as
\begin{equation}
\begin{aligned}
   D_k(\omega,u_2,u_3,E,\mu)
   &= \int dr\,ds\,e^{2iE(u_2 s+u_3 r)} \int dt\,e^{-i\omega t}\,
    \widetilde D_k(r,s,t,E,\mu) \,, \\
   D_k(\omega_1,\omega_2,u_2,E,\mu)
   &= \int ds\,e^{2iE u_2 s} \int dt\,du\,e^{-i(\omega_1 t+\omega_2 u)}\,
    \widetilde D_k(s,t,u,E,\mu) \,.
\end{aligned}
\end{equation}
In matching calculations, $u_3$ is identified with the longitudinal 
momentum fraction $x_3$ carried by a final-state collinear gluon, while
$\omega_1$ and $\omega_2$ are associated with the components $n\cdot l_q$ 
and $n\cdot l_g$ of the incoming soft anti-quark and gluon momenta. 

We are now in a position to construct a basis of four-quark operators 
relevant to the discussion of the universal functions $\zeta_M$ in 
(\ref{ff}). It is convenient {\em not\/} to use the relation 
$\vslash\,h=h$ in the matching from SCET$_{\rm I}$ onto SCET$_{\rm II}$, 
since then it is guaranteed that $\nslash$ must appear to the left of the 
heavy-quark field $\H$. (This follows from the Fierz relation 
(\ref{Fierz}) below, where in our case $M$ commutes with $\nslash$.) The 
Feynman rules of SCET imply that the resulting operators must contain an 
odd number of Dirac matrices besides the matrix $\Gamma_M$ contained in 
the effective current operator $J_M^{(0)}$. These operators must 
transform like the current $J_M^{(0)}$ under Lorentz transformations. 
Also, the soft and collinear parts of the four-quark operators must have 
non-zero projections onto the $B$ meson and the final-state meson $M$. We 
set the transverse momenta of the mesons to zero, in which case there is 
no need to include operators with transverse derivatives acting on the 
products of all soft or collinear fields.

\vspace{-0.2cm}
\paragraph{Case \boldmath$M=P$\unboldmath:\/}
The resulting operators must transform as a scalar ($\Gamma_M=1$). A 
basis of such operators is
\begin{equation}\label{Pbasis}
\begin{aligned}
   O_1^{(P)}
   &= g^2\,[\bar\X(0)\,\frac{\nbslash}{2}\,\gamma_5\,\X(s\bar n)]\,
    [\bar\Q_s(tn)\,\frac{\nbslash\nslash}{4}\,\gamma_5\,\H(0)] \,, \\
   O_2^{(P)}
   &= g^2\,[\bar\X(0)\,\frac{\nbslash}{2}\,\gamma_5\,
    i\delslash_\perp\X(s\bar n)]\,
    [\bar\Q_s(tn)\,\frac{\nslash}{2}\,\gamma_5\,\H(0)] \,, \\
   O_3^{(P)}
   &= g^2\,[\bar\X(0)\,\frac{\nbslash}{2}\,\gamma_5\,
    \calAslash_{c\perp}(r\bar n)\,\X(s\bar n)]\,
    [\bar\Q_s(tn)\,\frac{\nslash}{2}\,\gamma_5\,\H(0)] \,, \\
   O_4^{(P)}
   &= g^2\,[\bar\X(0)\,\frac{\nbslash}{2}\,\gamma_5\,\X(s\bar n)]\,
    [\bar\Q_s(tn)\,\calAslash_{s\perp}(un)\,\frac{\nslash}{2}\,\gamma_5\,
    \H(0)] \,.
\end{aligned}
\end{equation}
The soft and collinear currents both transform like a pseudoscalar.

\vspace{-0.2cm}
\paragraph{Case \boldmath$M=V_\parallel$:\unboldmath}
The resulting operators must transform as a pseudo-scalar 
($\Gamma_M=\gamma_5$). A basis of such operators is obtained by omitting 
the $\gamma_5$ between the two collinear spinor fields in (\ref{Pbasis}), 
so that the collinear currents transform like a scalar. The Wilson 
coefficients for the cases $M=P$ and $M=V_\parallel$ coincide up to an 
overall sign due to parity invariance.

\vspace{-0.2cm}
\paragraph{Case \boldmath$M=V_\perp$:\unboldmath}
The resulting operators must transform as a transverse vector 
($\Gamma_M=\gamma_{\perp\nu}$). A basis of such operators is
\begin{equation}
\begin{aligned}
   O_1^{(V_\perp)}
   &= g^2\,[\bar\X(0)\,\frac{\nbslash}{2}\,
    \gamma_{\perp\nu}\gamma_5\,\X(s\bar n)]\,
    [\bar\Q_s(tn)\,\frac{\nbslash\nslash}{4}\,\gamma_5\,\H(0)] \,, \\
   O_2^{(V_\perp)}
   &= g^2\,[\bar\X(0)\,\frac{\nbslash}{2}\,
    (i\epsilon_{\nu\alpha}^\perp - g_{\nu\alpha}^\perp\gamma_5)\,
    i\partial_\perp^\alpha\,\X(s\bar n)]\,
    [\bar\Q_s(tn)\,\frac{\nslash}{2}\,\gamma_5\,\H(0)] \,, \\
   O_3^{(V_\perp)}
   &= g^2\,[\bar\X(0)\,\frac{\nbslash}{2}\,
    (i\epsilon_{\nu\alpha}^\perp - g_{\nu\alpha}^\perp\gamma_5)\,
    \A_{c\perp}^\alpha(r\bar n)\,\X(s\bar n)]\,
    [\bar\Q_s(tn)\,\frac{\nslash}{2}\,\gamma_5\,\H(0)] \,, \\
   O_4^{(V_\perp)}
   &= g^2\,[\bar\X(0)\,\frac{\nbslash}{2}\,
    \gamma_{\perp\nu}\gamma_5\,\X(s\bar n)]\,
    [\bar\Q_s(tn)\,\calAslash_{s\perp}(un)\,\frac{\nslash}{2}\,
    \gamma_5\,\H(0)] \,.
\end{aligned}
\end{equation}
The soft currents transform like a pseudoscalar, while the collinear
currents transform like a transverse axial-vector. (The relative sign
between the two terms in $O_{2,3}^{(V_\perp)}$ can be determined by
considering left and right-handed spinors $\bar\X$ and using relation
(\ref{wonderful}) below.)\\

\noindent
It is not necessary to include operators with a derivative on the soft 
spectator anti-quark. All such operators would contain 
$\bar Q_s\,(-i\overleftarrow{\delslash}_{\!\!\perp})\,\nslash$, which can 
be related to a linear combination of the operators $O_1^{(M)}$ and 
$O_4^{(M)}$ using the equation of motion for the light-quark field. In 
our definitions above we have included the QCD coupling constant into the 
operators rather than the Wilson coefficient functions. This is natural, 
since a factor $g^2$ (not $\alpha_s/\pi$) already arises in tree-level 
matching. More importantly, as we will see the matrix elements of the 
operators $O_i^{(M)}$ are {\em not\/} dominated by physics at the 
hard-collinear scale, but rather by long-distance hadronic physics.
The coupling should therefore be evaluated at a low scale 
$\mu\ll\sqrt{E\Lambda}$.

\subsubsection{Wilson coefficients of four-quark operators}

We now calculate the momentum-space Wilson coefficients of the operators 
$O_i^{(M)}$ at tree level. The relevant graphs must contain a 
hard-collinear gluon exchange, which turns the soft $B$-meson spectator 
anti-quark into a collinear parton that can be absorbed by the 
final-state hadron. These hard-collinear gluons are integrated out when 
SCET$_{\rm I}$ is matched onto SCET$_{\rm II}$. However, at $O(\alpha_s)$
we can obtain the Wilson coefficients in SCET$_{\rm II}$ by directly 
matching QCD amplitudes onto the low-energy theory, without going through 
an intermediate effective theory.

As we have seen, at leading power the universal form factors $\zeta_M$ 
receive contributions from ordinary four-quark operators as well as from 
four-quark operators containing an additional collinear or soft gluon 
field. We start with a discussion of the matching calculation for the
operators $O_{1,2}^{(M)}$, whose coefficients can be obtained by 
analyzing the diagrams shown in Figure~\ref{fig:twopart} in the kinematic 
region where the outgoing quarks are collinear and the incoming quarks 
are soft. The resulting expression for the amplitude in the full theory 
has already been given in (\ref{LO}). Note that, by assumption, 
$n\cdot l\sim\Lambda$ and $x_2\sim 1$, so that the result is well defined 
and it is consistent to neglect subleading terms in the gluon 
propagators. 

We wish to express the result for the QCD amplitude as a combination of 
SCET$_{\rm II}$ matrix elements multiplied by coefficient functions. To 
this end, we must first eliminate the superficially leading term in 
(\ref{LO}) using the equations of motion. Assigning momenta to the two 
outgoing collinear lines as shown in (\ref{p1p2}), it follows that
\begin{equation}\label{eom}
   \gamma_\mu E\nslash\,\Gamma * \gamma^\mu
   = \frac{\pslash_\perp}{x_1}\,\gamma_\mu\Gamma * \gamma^\mu
   - 2\Gamma * \frac{\pslash_\perp}{x_2} + \dots \,,
\end{equation}
where the dots denote power-suppressed terms. In the next step we use a 
Fierz transformation to recast the amplitude into a form that is 
convenient for our analysis. Taking into account that between collinear 
spinors the Dirac basis contains only three independent matrices, we 
find for general matrices $M$ and $N$
\begin{eqnarray}\label{Fierz}
  2 (\bar u_\xi M u_h)\,(\bar v_q N v_\xi) 
  &=& (\bar u_\xi \frac{\nbslash}{2} v_\xi)\,
   (\bar v_q N \frac{\nslash}{2} M u_h) 
   + (\bar u_\xi \frac{\nbslash}{2}\gamma_5 v_\xi)\,
   (\bar v_q N \gamma_5 \frac{\nslash}{2} M u_h) \nonumber \\
  &&\mbox{}+ (\bar u_\xi \frac{\nbslash}{2}\gamma_{\perp\alpha} v_\xi)\,
   (\bar v_q N \gamma_\perp^\alpha \frac{\nslash}{2} M u_h) \,.
\end{eqnarray}
To simplify the Dirac algebra we use the identities
\begin{equation}\label{wonderful}
   \gamma_\perp^\mu\gamma_5\,\nslash
    = i\epsilon_\perp^{\mu\nu}\gamma_{\perp\nu}\,\nslash \,, \qquad
   \gamma_\perp^\mu\gamma_\perp^\nu\,\nslash
    = (g_\perp^{\mu\nu} - i\epsilon_\perp^{\mu\nu}\gamma_5)\,\nslash \,.
\end{equation}
Finally, we include a minus sign from fermion exchange under the Fierz
transformation and project the collinear and soft ``currents'' in the 
resulting expressions onto color-singlet states.

Once in this form, the different contributions to the amplitude can be
readily identified with matrix elements of the operators $O_{1,2}^{(M)}$.
For $M=P,V_\parallel$ we obtain
\begin{equation}
   D_1^{(P)} = - D_1^{(V_\parallel)}
    = - \frac{C_F}{N}\,\frac{1+u_2}{4E^2 u_2^2\,\omega} \,, \qquad
   D_2^{(P)} = - D_2^{(V_\parallel)}
    = - \frac{C_F}{N}\,\frac{1}{4E^2 u_1 u_2^2\,\omega^2} \,,
\end{equation}
while for $M=V_\perp$ we find
\begin{equation}
   D_1^{(V_\perp)}
    = \frac{C_F}{N}\,\frac{1}{4E^2 u_2^2\,\omega} \,, \qquad
   D_2^{(V_\perp)} = \frac{C_F}{N}\,\frac{1}{4E^2 u_2^2\,\omega^2} \,.
\end{equation}

\subsubsection{Wilson coefficients of four-quark operators with an extra 
gluon}

\begin{figure}
\begin{center}
\epsfig{file=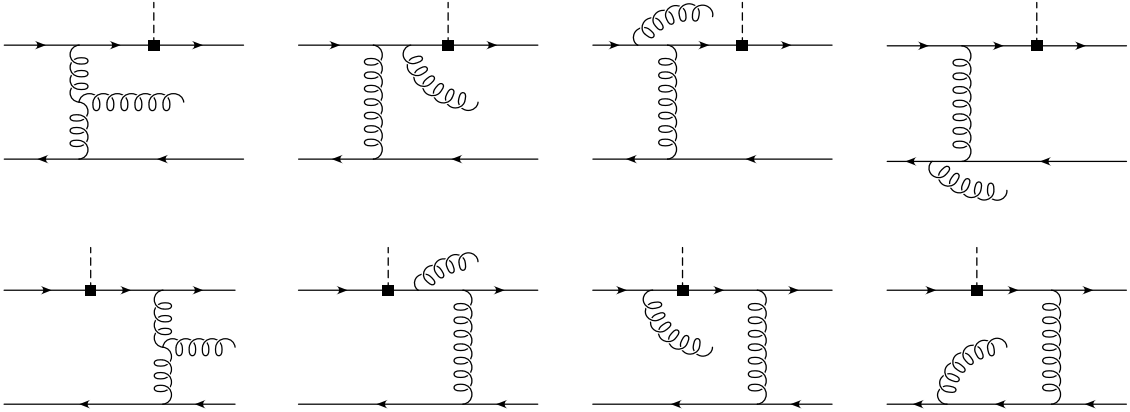,width=15cm}
\end{center}
\centerline{\parbox{14cm}{\caption{\label{fig:1}
Diagrams relevant to the matching calculation for operators containing an 
extra collinear gluon.}}}
\end{figure}

The matching calculation for the operators $O_{3,4}^{(M)}$ proceeds along 
the same lines. However, in this case it is necessary to study diagrams 
with four external quarks and an external gluon, which we treat as a 
background field. Let us first discuss the case with three collinear 
particles in the final state. The relevant QCD graphs are shown in 
Figure~\ref{fig:1}. Physically, they correspond to non-valence Fock 
states of the light final-state meson, but this interpretation is 
irrelevant for the matching calculation, which can be done with free 
quark and gluon states. 

By assumption, each of the three collinear particles carries large 
momentum components along the $n$ direction. We denote the corresponding 
longitudinal momentum fractions by $x_1$ (quark), $x_2$ (anti-quark) and
$x_3$ (gluon), where $x_1+x_2+x_3=1$. The calculation of the diagrams in 
Figure~\ref{fig:1} exhibits that at leading power the amplitude depends 
only on the light-cone components $n\cdot l$ and $\bar n\cdot p_i=2E x_i$ 
of the external momenta, and that only transverse components of the 
external gluons fields must be kept. These observations are in accordance 
with the structure of the operator $O_3^{(M)}$. One must add to the 
diagrams shown in the figure a contribution arising from the application 
of the equation of motion for the collinear quark fields that led to 
(\ref{eom}). One way of obtaining it is to include diagrams with gluon 
emission from the external collinear quark lines in the matching 
calculation. Performing the Fierz transformation and projecting onto the 
color singlet-singlet operators $O_3^{(M)}$, we obtain the coefficient 
functions
\begin{equation}
\begin{aligned}
   D_3^{(P)} = - D_3^{(V_\parallel)}
   &= \frac{1}{8E^2(u_2+u_3)^2\omega^2} \left[
    \frac{u_3}{u_2} - 1 + \frac{2}{N^2} - \frac{2C_F}{N}
    \frac{(u_2+u_3)^2}{u_2(u_1+u_3)} \right] , \\
   D_3^{(V_\perp)}
   &= - \frac{1}{8E^2(u_2+u_3)^2\omega^2} \left[
    \frac{u_3}{u_2} - 1 + \frac{2}{N^2}
    + \frac{1}{N^2}\,\frac{(u_2+u_3)^2}{u_2(u_1+u_2)} \right] .
\end{aligned}
\end{equation}

Next, we calculate the contributions from diagrams with three external 
soft particles, which physically correspond to three-particle Fock states
of the $B$ meson. Again there are eight diagrams, analogous to those 
in Figure~\ref{fig:1}. The initial soft gluon is attached to either an 
off-shell intermediate line or a collinear line. The diagrams with 
external gluons must again be supplemented by a contribution resulting 
from the application of the equation of motion $\bar v_q\,\lslash=O(g)$ 
used in the analysis of the four-quark amplitude in (\ref{LO}). After 
Fierz transformation and projection onto the color singlet-singlet 
operators $O_4^{(M)}$, we find the coefficient functions
\begin{equation}
\begin{aligned}
   D_4^{(P)} = - D_4^{(V_\parallel)}
   &= \frac{1}{8E^2 u_2^2(\omega_1+\omega_2)^2} \left[
    \left( 1 - \frac{2C_F}{N}\,u_2 \right) 
    \frac{\omega_2}{\omega_1} + \frac{1}{N^2} \right] , \\
   D_4^{(V_\perp)} 
   &= - \frac{1}{8E^2 u_2^2(\omega_1+\omega_2)^2} \left[
    \left( 1 + \frac{1}{N^2}\,\frac{u_2}{u_1} \right)
    \frac{\omega_2}{\omega_1}
    + \frac{1}{N^2}\,\frac{1}{u_1} \right] .
\end{aligned}
\end{equation}
The variables $\omega_1$ and $\omega_2$ correspond to the light-cone
components $n\cdot l_q$ and $n\cdot l_g$ of the incoming momenta of the 
soft anti-quark and gluon, respectively.

\subsubsection{Endpoint singularities}
\label{subsec:divs}

The Wilson coefficients of the SCET$_{\rm II}$ four-quark operators 
become singular in the limit where some of the momentum components of the 
external particles tend to zero. When the $B\to M$ matrix elements of the 
operators $O_k^{(M)}$ are evaluated, these singularities give rise to
endpoint divergences of the resulting convolution integrals with LCDAs.
For instance, the matrix elements of the operators $O_1^{(M)}$ and 
$O_4^{(M)}$ involve the leading twist-2 projection onto the light meson
$M$. The corresponding LCDAs $\phi_M$ are expected to vanish linearly as 
$u_2\to 0$, whereas the corresponding coefficients contain terms that 
grow like $1/u_2^2$, giving rise to logarithmically divergent convolution 
integrals. Similarly, the operators $O_2^{(M)}$ and $O_3^{(M)}$ involve 
the leading-order projection onto the $B$-meson LCDA $\phi_B^{(+)}$, 
which is expected to vanish linearly as $\omega\to 0$ 
\cite{Lange:2003ff}. Once again, logarithmic singularities arise because 
the corresponding coefficients contain terms that grow like $1/\omega^2$. 

While the logarithmic divergences in the convolution integrals could be 
avoided by introducing some infra-red regulators, they indicate that 
leading-order contributions to the amplitudes arise from momentum regions 
that cannot be described correctly in terms of collinear or soft fields. 
In Section~\ref{sec:toy}, we will explain that in SCET$_{\rm II}$ these 
configurations are accounted for by matrix elements of operators 
containing the soft-collinear messenger fields.

It is interesting to note that the operators $O_{3,4}^{(M)}$ arise from 
the matching of the SCET$_{\rm I}$ operator called $T_0^{\rm F}$ in
\cite{Bauer:2002aj} onto operators of the low-energy effective theory. 
The fact that the matrix elements of these operators contain endpoint 
divergences shows that the criterion of hard-collinear/soft factorization 
of SCET$_{\rm I}$ operators is insufficient to decide whether matrix 
elements can be factorized into separate soft and collinear entities in 
SCET$_{\rm II}$ or not. 

\subsection{Spin-symmetry breaking contributions}

The two-particle amplitude in (\ref{LO}) also includes contributions for 
which the Dirac structure is different from $\nslash\,\Gamma\,h$. These 
give rise to symmetry-breaking contributions to the form factors. We 
shall not derive a complete basis of all possible symmetry-breaking 
operators (there are many) but rather list the ones that enter at first 
order in $\alpha_s$. Since the spin-symmetry violating terms can be 
factorized in the form of the second term in (\ref{ff}), they are 
associated with a short-distance coupling constant 
$\alpha_s(\sqrt{E\Lambda})$. It is therefore appropriate in this case to 
include the coupling constant in the Wilson coefficient functions.

Let $\Gamma$ denote the Dirac structure of the flavor-changing currents
$\bar q\,\Gamma\,b$, whose matrix elements define the form factors. To 
first order in $\alpha_s$, the spin-symmetry breaking terms can then be 
obtained from the matrix elements of two operators given by
\begin{equation}\label{Q1Q2}
\begin{aligned}
   Q_1 &= [\bar\X(0)\,\frac{\nbslash}{2} \left(
    \begin{array}{c} 1 \\ \gamma_5 \\ \gamma_{\perp\alpha} \end{array}
    \right) \X(s\bar n)]\,
    [\bar\Q_s(tn)\,\frac{\nslash}{2}\,\gamma_\mu\! \left( 
    \begin{array}{c} 1 \\ -\gamma_5 \\ -\gamma_\perp^\alpha \end{array}
    \right) \Gamma\,\gamma^\mu\nslash\,\H(0)] \,, \\
   Q_2 &= [\bar\X(0)\,\frac{\nbslash}{2} \left(
    \begin{array}{c} 1 \\ \gamma_5 \\ 0 \end{array} \right) \X(s\bar n)]\,
    [\bar\Q_s(tn)\,\frac{\nslash\nbslash}{4}
    \left( \begin{array}{c} 1 \\ -\gamma_5 \\ 0 \end{array} \right)
    \Gamma\,\H(0)] \,,
\end{aligned}
\end{equation}
where each line contributes for a different final-state meson $M$. The 
corresponding Wilson coefficients are
\begin{equation}
   \hat T_1 = - \frac{C_F}{N}\,\frac{\pi\alpha_s}{2m_B E u_2\,\omega} \,,
    \qquad
   \hat T_2 = \frac{C_F}{N}\,\frac{\pi\alpha_s}{E^2 u_2\,\omega} \,.
\end{equation}
Linear combinations of $\hat T_1$ and $\hat T_2$ determine (up to 
prefactors) the hard-scattering kernels $T_i$ in (\ref{ff}). The matrix 
elements of the operators $Q_{1,2}$ can be expressed in terms of the 
leading-order LCDAs of the $B$ meson and the light meson $M$. Only the 
$B$-meson LCDA called $\phi_B^{(+)}$ contributes \cite{Grozin:1996pq} 
because of the factor $\nslash$ next to $\bar Q_s$. This property holds
true to all orders in perturbation theory \cite{Bauer:2002aj}. The 
resulting convolution integrals are convergent. Evaluating these matrix 
elements we reproduce the spin-symmetry breaking terms obtained in 
\cite{Beneke:2000wa}.

\section{Physics of endpoint singularities -- a toy model}
\label{sec:toy}

Our strategy in the previous sections has been to perform matching 
calculations by expanding QCD amplitudes in powers of $\Lambda/E$, 
assuming that collinear and soft external momenta have the scaling 
assigned to them in SCET. We have then matched the results onto 
SCET$_{\rm II}$ operators containing soft and collinear fields and read 
off the corresponding Wilson coefficient functions in momentum space. A 
problematic aspect of this procedure has been the observation that, if 
the matrix elements of the effective-theory operators are expressed in
terms of meson LCDAs, then the resulting convolution integrals do not 
converge. Endpoint singularities arise, which correspond to exceptional 
momentum configurations in which some of the partons inside the external
hadrons carry very small momentum. The question naturally arises how one 
should interpret these singularities, and whether the results we found 
for the short-distance coefficient functions are in fact correct.

\begin{figure}
\begin{center}
\epsfig{file=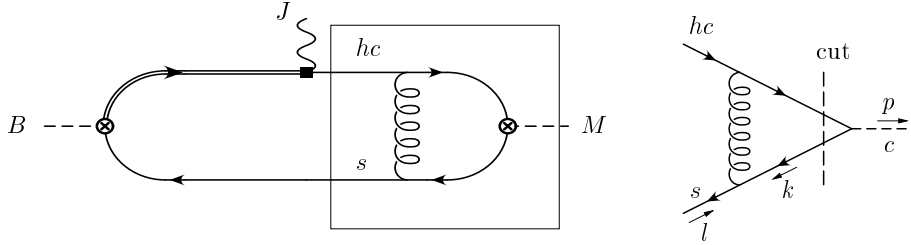,width=12cm} 
\end{center}
\centerline{\parbox{14cm}{\caption{\label{fig:toy}
Triangle subgraph whose spectral function can be used to study endpoint 
singularities on the collinear side.}}}
\end{figure}

The fact that endpoint configurations are not kinematically suppressed 
points to the relevance of new momentum modes. In the limit $x_2\to 0$, 
the scaling associated with the collinear anti-quark in the final state
of Figure~\ref{fig:twopart} changes from $(\lambda^2,1,\lambda)$ to
$(\lambda^2,\lambda,\dots)$. Likewise, in the limit $n\cdot l\to 0$, the 
scaling associated with the soft spectator anti-quark in the initial 
state changes from $(\lambda,\lambda,\lambda)$ to 
$(\lambda^2,\lambda,\dots)$. The soft-collinear messenger fields in the
SCET$_{\rm II}$ Lagrangian have precisely the scaling properties 
corresponding to these exceptional configurations. (Since the modes in an 
effective theory are always on-shell, the transverse components of 
soft-collinear momenta scale like $\lambda^{3/2}$. We will see below that 
this is not really relevant.)

The purpose of this section is to analyze the interplay of the various 
modes present in SCET$_{\rm II}$ and to see how soft-collinear messengers 
are connected with the phenomenon of endpoint singularities. We will do 
this with the help of a toy example. Consider the triangle subgraph 
enclosed by the box in the diagram shown on the left-hand side in 
Figure~\ref{fig:toy}, which is one of the two gluon-exchange graphs 
relevant to the $B\to M$ form factors (see Figure~\ref{fig:twopart}). 
To understand the physics of endpoint singularities we focus on the side 
of the light meson $M$ (an analogous discussion could be given for the 
$B$-meson side). We study the discontinuity of the triangle diagram in 
the external collinear momentum $p^2$. The resulting spectral density 
$\varrho(p^2)$ models the continuum of light final-state hadrons that can 
be produced on the collinear side. We will not bother to project out a 
particular light meson from this spectral density. For simplicity, we 
will also ignore any numerator structure in the triangle subgraph and 
instead study the corresponding scalar triangle, which has already been 
discussed in \cite{Becher:2003qh}. The results we obtain are nevertheless 
general.

Let us define the discontinuity
\begin{equation}
   \frac{1}{2\pi i}\,\Big[ I(p^2+i0) - I(p^2-i0) \Big]
   \equiv D\cdot\theta(p^2)
\end{equation}
of the scalar triangle integral
\begin{equation}
   I = i\pi^{-d/2} \mu^{4-d} \int d^dk\,
   \frac{2l_+\cdot p_-}{(k^2+i0)\,[(k+l)^2+i0]\,[(k+p)^2+i0]}
\end{equation}
in $d=4-2\epsilon$ space-time dimensions, where $p$ is the outgoing 
collinear momentum and $l$ the incoming soft momentum. It will be 
convenient to define the invariants $L^2\equiv-l^2-i0$ and 
$Q^2\equiv-(l-p)^2=2l_+\cdot p_-+\dots$, which scale like 
$L^2\sim\lambda^2$ and $Q^2\sim\lambda$. (In physical units, 
$L^2\sim\Lambda^2$ and $Q^2\sim E\Lambda$ with $E\gg\Lambda$.) We will 
assume that these quantities are non-zero. From \cite{Becher:2003qh}, we 
can obtain explicit results for the discontinuity $D$ and for the various 
regions of loop momentum $k$ that give a non-vanishing contribution. We 
find that 
\begin{equation}
   D = \ln\frac{Q^2}{L^2} + O(\epsilon,\lambda) \,,
\end{equation}
and that the momentum configurations that contribute to this result are 
those where the loop momentum is either collinear, meaning that
$k\sim(\lambda^2,1,\lambda)$, or soft-collinear, meaning that
$k\sim(\lambda^2,\lambda,\lambda^{3/2})$. The contributions from these 
two regions are
\begin{equation}\label{DcDsc}
\begin{aligned}
   D_{\rm C} &= \frac{\Gamma(-\epsilon)}{\Gamma(1-2\epsilon)}
    \left( \frac{\mu^2}{p^2} \right)^\epsilon
    = - \frac{1}{\epsilon} + \gamma_E - \ln\frac{\mu^2}{p^2}
    + O(\epsilon) \,, \\
   D_{\rm SC} &= \Gamma(\epsilon)
    \left( \frac{\mu^2 Q^2}{p^2 L^2} \right)^\epsilon
    = \frac{1}{\epsilon} - \gamma_E + \ln\frac{\mu^2 Q^2}{p^2 L^2}
    + O(\epsilon) \,,
\end{aligned}
\end{equation}
which add up to the correct answer. 

It is instructive to rewrite these results in a more transparent form. To 
this end we use Cutkosky rules to evaluate the discontinuities of the 
diagrams directly and perform all phase-space integrations except the 
integral over the light-cone component $\bar n\cdot k$ of the loop 
momentum, which we parameterize as 
$\bar n\cdot k\equiv-x_2\bar n\cdot p$. As in previous sections, $x_2$ 
denotes the fraction of longitudinal momentum carried by the anti-quark 
in the final-state. The exact result for the discontinuity of the scalar
triangle is ($\bar x_2\equiv 1-x_2$)
\begin{eqnarray}\label{Dfull}
   D &=& \frac{1}{\Gamma(1-\epsilon)}
    \left( \frac{\mu^2}{p^2} \right)^\epsilon
    \int_0^1\!dx_2\,\frac{(x_2\bar x_2)^{-\epsilon}}{x_2+L^2/Q^2}
    + O(\lambda) \nonumber\\
   &=& \int_0^1\!dx_2\,\frac{1}{x_2+L^2/Q^2} + O(\epsilon,\lambda)
    = \int_{L^2/Q^2}^1 \frac{dx_2}{x_2} + O(\epsilon,\lambda) \,.
\end{eqnarray}
The collinear and soft-collinear contributions separately are divergent 
even though they correspond to tree diagrams (after the two propagators 
have been cut). We obtain
\begin{equation}\label{DcDsc2}
\begin{aligned}
   D_{\rm C} &= \frac{1}{\Gamma(1-\epsilon)}
    \left( \frac{\mu^2}{p^2} \right)^\epsilon
    \int_0^1 \frac{dx_2}{x_2}\,(x_2\bar x_2)^{-\epsilon} \,, \\
   D_{\rm SC} &= \frac{1}{\Gamma(1-\epsilon)}
    \left( \frac{\mu^2}{p^2} \right)^\epsilon
    \int_0^\infty\!dx_2\,\frac{x_2^{-\epsilon}}{x_2+L^2/Q^2}
    \,.
\end{aligned}
\end{equation}
The collinear contribution is infra-red singular for $x_2\to 0$, which is 
an example of an endpoint singularity. In the present case, the 
singularity is regularized dimensionally by keeping $\epsilon$ non-zero. 
The soft-collinear contribution is infra-red finite but ultra-violet 
divergent, since $x_2$ runs up to $\infty$. Again, this divergence is 
regularized dimensionally. Evaluating the remaining integrals one 
recovers the exact results in (\ref{DcDsc}). 

\begin{figure}
\begin{center}
\epsfig{file=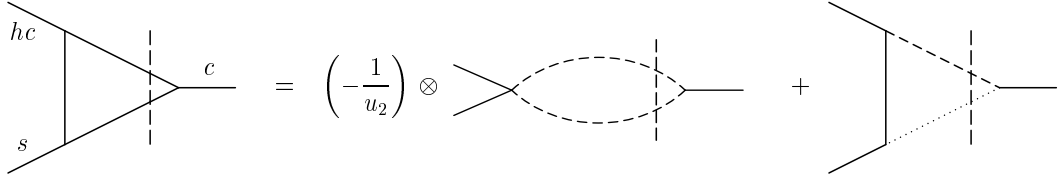,width=14cm} 
\end{center}
\centerline{\parbox{14cm}{\caption{\label{fig:toy2}
Collinear and soft-collinear contributions to the scalar triangle graph 
in SCET$_{\rm II}$. Dashed (dotted) lines represent collinear 
(soft-collinear) propagators.}}}
\end{figure}

Several comments are in order:

i)\;
The result for the spectral density in the full theory is finite. The
endpoint singularity is regularized by keeping subleading terms in the
hard-collinear propagator $1/[-(k+l)^2]\simeq 1/(x_2 Q^2+L^2)$. When the 
subleading terms are dropped based on naive power counting (as we did in 
the analysis of the previous sections) the full-theory result reduces to 
the contribution obtained from the collinear region. An endpoint 
singularity arises in this case, which however can be regularized 
dimensionally. In order for this to happen in the realistic case of
external meson states, it would be necessary to perform the projections 
onto the meson LCDAs in $d\ne 4$ dimensions. The factor 
$(x_2\bar x_2)^{-\epsilon}$ in (\ref{DcDsc2}) would then correspond to a
modification of the LCDAs, which renders the convolution integrals 
finite.

ii)\;
The collinear approximation fails for values $x_2=O(\lambda)$. For such 
small momentum fractions the exact full-theory result is reproduced by 
the contribution obtained from the soft-collinear region. In fact, the 
collinear and soft-collinear contributions in (\ref{DcDsc2}) coincide 
with the first terms in the Taylor expansion of the full-theory result in 
(\ref{Dfull}) in the limits where $x_2=O(1)\gg L^2/Q^2$ and 
$x_2=O(\lambda)\ll 1$. The fact that $x_2$ runs up to $\infty$ in the 
soft-collinear contribution is not a problem. In dimensional 
regularization the integral receives significant contributions only from 
the region where $x_2\sim L^2/Q^2=O(\lambda)$. To see this, one can 
introduce a cutoff $\delta$ to separate the collinear and soft-collinear 
contributions, chosen such that $1\gg\delta\gg\lambda$. This cutoff is 
introduced as an infra-red regulator in the collinear integral 
($\int_0^1\to\int_\delta^1$) and as an ultra-violet regulator in the 
soft-collinear integral ($\int_0^\infty\to\int_0^\delta$). The integrals 
can then be evaluated setting $\epsilon\to 0$. This yields 
$D_{\rm C}=-\ln\delta$ and 
$D_{\rm SC}=\ln\delta+\ln\frac{Q^2}{L^2}+O(\lambda/\delta)$. The sum of 
the two contributions once again reproduces the exact result.

iii)\; 
Next, note that the soft-collinear contribution {\em precisely\/} 
reproduces the endpoint behavior of the full-theory amplitude. It is 
irrelevant in this context that the soft-collinear virtuality 
$k_{sc}^2\sim E^2\lambda^3$ is parametrically smaller than the QCD scale 
$\Lambda^2$. What matters is that the plus and minus components of the
soft-collinear momentum, $k_{sc}\sim(\lambda^2,\lambda,\dots)$, are of 
the same order as the corresponding components of a collinear momentum in 
the endpoint region, where $\bar n\cdot p\sim\lambda$ rather than being 
$O(1)$.

iv)\;
Finally, we see that the endpoint divergences we encountered in the 
Section~\ref{subsec:divs} were not regularized because we dropped the 
dimensional regulator when performing the projections onto meson LCDAs. 
This, however, does not affect the results for the Wilson coefficients of 
the SCET$_{\rm II}$ operators, which remain valid. In the toy model, the 
collinear contribution is represented in the effective theory as the 
discontinuity of the integral (corresponding to the first diagram on the 
right-hand side in Figure~\ref{fig:toy2})
\begin{eqnarray}
   I_{\rm C}
   &=& i\pi^{-d/2} \mu^{4-d} \int d^dk\,
    \frac{2l_+\cdot p_-}{(k^2+i0)\,(2k_-\cdot l_+)\,[(k+p)^2+i0]}
    \nonumber\\
   &=& i\pi^{-d/2} \mu^{4-d} \int d^dk\,\frac{(-1)}{x_2}\,
    \frac{1}{(k^2+i0)\,[(k+p)^2+i0]} \,,
\end{eqnarray}
which shows that the Wilson coefficient resulting from integrating out 
the hard-collinear propagator is simply $-1/u_2$.

To summarize this discussion, we stress that to reproduce the behavior of 
the full amplitude it is necessary to include in the effective theory 
contributions involving collinear fields and those involving 
soft-collinear messengers, as indicated in Figure~\ref{fig:toy2}. The 
latter ones represent the exact behavior of the amplitude in the endpoint 
region. Whether or not endpoint configurations contribute at leading 
order in power counting is equivalent to the question of whether or not 
operators involving soft-collinear fields can arise at leading power. 
This shows the power of our formalism. Operators containing 
soft-collinear fields can be used to parameterize in a systematic way the 
long-distance physics associated with endpoint configurations in the 
external mesons states. This will be discussed in more detail in the next 
section. It also follows that the scaling properties of operators 
containing soft-collinear fields can be used to make model-independent 
statements about the convergence of convolution integrals that arise in 
QCD factorization theorems such as (\ref{ff}). We will exploit this 
connection in Section~\ref{sec:fact}.

\section{Soft-collinear messengers and the soft overlap}
\label{sec:softoverlap}

In Section~\ref{sec:matching} we have derived four-quark operators built 
out of collinear and soft fields, which contribute at leading order to 
the universal soft functions $\zeta_M$ in (\ref{ff}). Power counting 
shows that the products of these operators with their Wilson coefficients 
scale like $\lambda^4$. (Here one uses that $dr\sim ds\sim 1$ and 
$dt\sim du\sim\lambda^{-1}$.) Taking into account the counting of the 
external hadron states, $|B\rangle\sim\lambda^{-3/2}$ and 
$\langle M|\sim\lambda^{-1}$, it follows that the corresponding 
contributions to the universal form factors scale like $\lambda^{3/2}$. 
In other words, heavy-to-light form factors are quantities that vanish at 
leading order in the large-energy expansion. The fact that the resulting 
convolution integrals were infra-red singular suggested that there should 
be other contributions of the same order in power counting, which cannot 
be described in terms of collinear or soft fields.

An important observation made in the previous section was that the 
endpoint behavior of QCD amplitudes is described in the low-energy theory 
in terms of diagrams involving soft-collinear messenger fields. For 
instance, in the last graph in Figure~\ref{fig:toy2} the soft spectator 
anti-quark inside the $B$ meson turns into a soft-collinear anti-quark by 
the emission of a soft gluon. The soft-collinear anti-quark is then 
absorbed by the final-state meson. In a similar way, endpoint 
singularities on the $B$-meson side would correspond to a situation where 
the initial state contains a soft-collinear spectator anti-quark, which 
absorbs a collinear gluon and turns into a collinear anti-quark. The 
SCET$_{\rm II}$ Lagrangian contains the corresponding interaction terms 
only at subleading order in $\lambda$. However, because the universal 
form factors $\zeta_M$ themselves are power-suppressed quantities, these 
subleading interactions will nevertheless give rise to leading-power 
contributions.

We will need the first two non-vanishing orders in the interactions that 
couple a soft-collinear quark to a soft or collinear quark. The results 
are most transparent when expressed in terms of the gauge-invariant 
building blocks introduced in (\ref{blocks}). They are 
\cite{Becher:2003qh}
\begin{equation}
   {\cal L}_{\bar q\theta}^{(1/2)}
   = \bar\Q_s\,\calAslash_{s\perp} W_{sc}^\dagger\,\theta \,, \qquad 
   {\cal L}_{\bar\theta\xi}^{(1/2)}
   = \bar\sigma\,S_{sc}\,\calAslash_{c\perp}\,\X \,, \vspace{-0.2cm}
\end{equation}
and
\begin{equation}
\begin{aligned}
   {\cal L}_{\bar q\theta}^{(1)}
   &= \bar\Q_s\,\calAslash_{s\perp} W_{sc}^\dagger\,
    (x_\perp\cdot D_{sc}\,\theta + \sigma) 
    + \bar\Q_s\,\frac{\nslash}{2}\,\bar n\cdot\A_s\,
    W_{sc}^\dagger\,\sigma \,, \\
   {\cal L}_{\bar\theta\xi}^{(1)}
   &= \bar\sigma\,x_\perp\cdot\overleftarrow{D}_{sc}\,S_{sc}\,
    \calAslash_{c\perp}\,\X
    - \bar\theta\, S_{sc}\,\calAslash_{c\perp}\,\frac{\nbslash}{2}\,
    \frac{1}{i\bar n\cdot\partial}\,
    (i\delslash_\perp + \calAslash_{c\perp})\,\X
    + \bar\theta\,S_{sc}\,\frac{\nbslash}{2}\,n\cdot\A_c\,\X \,,
\end{aligned}
\end{equation}
where 
\begin{equation}\label{sigmasol}
   \sigma = - \frac{\nbslash}{2}\,\frac{1}{i\bar n\cdot D_{sc}}\,
   i\Dslash_{sc\perp}\,\theta
\end{equation}
contains the ``small components'' of the soft-collinear quark field, 
which are integrated out in the construction of the effective Lagrangian.
The longitudinal components of the calligraphic gluon fields are defined 
in \cite{Becher:2003qh}. (Note also that $n\cdot\A_s=0$ and 
$\bar n\cdot\A_c=0$.) The soft-collinear fields enter these results in 
combinations such as $W_{sc}^\dagger\,\theta$ or 
$S_{sc}^\dagger\,\theta$, which are gauge invariant. Soft and collinear 
fields live at position $x$, while soft-collinear fields are evaluated at 
position $x_+$ for ${\cal L}_{\bar q\theta}$ and $x_-$ for 
${\cal L}_{\bar\theta\xi}$. The measure $d^4x$ associated with these 
interactions scales like $\lambda^{-4}$. The superscript on the 
Lagrangians indicates at which order in power counting ($\lambda^{1/2}$ 
or $\lambda$) the corresponding terms contribute to the action.

Next, we need the representation of the flavor-changing SCET$_{\rm I}$ 
current $J_M^{(0)}$ in (\ref{current}) in terms of operators in the 
low-energy theory SCET$_{\rm II}$. As shown in
\cite{Becher:2003kh}, at leading power, and at the matching scale 
$\mu=\mu_{hc}$, the relation reads
\begin{equation}\label{Jmatch}
   J_M^{(0)}(x) 
   \to {\cal J}_M^{(0)}(x) = \bar\X(x_+ + x_\perp)\,\Gamma_M\,
   (S_{sc}^\dagger\,W_{sc})(0)\,\H(x_- + x_\perp)
\end{equation}
with a Wilson coefficient equal to unity. The anomalous dimensions of the 
currents are the same in the two theories. The product 
$(S_{sc}^\dagger\,W_{sc})(0)$ arises since soft-collinear messenger 
fields cannot be decoupled from the current operator ${\cal J}_M^{(0)}$ 
in SCET$_{\rm II}$, contrary to the case of the color singlet-singlet 
four-quark operators discussed in Section~\ref{sec:matching}.

\begin{figure}
\begin{center}
\epsfig{file=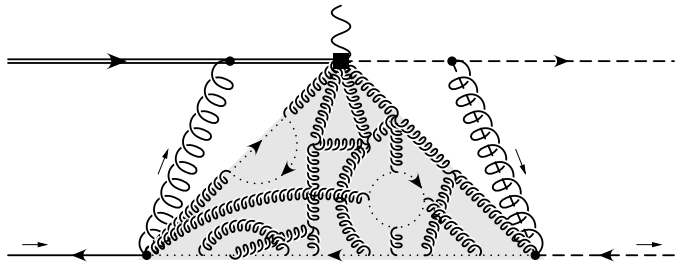,width=9cm} 
\end{center}
\centerline{\parbox{14cm}{\caption{\label{fig:SCblob}
An artist's view of the soft-collinear messenger contribution to the form
factors. The shaded region contains soft-collinear interactions. The 
arrows indicate the flow of the components $n\cdot p_s$ (left) and 
$\bar n\cdot p_c$ (right) of soft and collinear momenta, which do not 
enter the soft-collinear block.}}}
\end{figure}

Using these results, we can write down a tri-local operator whose matrix 
element provides a long-distance contribution to the universal form 
factors. It is
\begin{equation}\label{3T}
   O_5^{(M)} = i^2\!\int d^4x\,d^4y\,\mbox{T} \left\{
    {\cal L}_{\bar q\theta}^{(1/2)}(x)\,
    {\cal L}_{\bar\theta\xi}^{(1)}(y)\,{\cal J}_M^{(0)}(0)
    + {\cal L}_{\bar q\theta}^{(1)}(x)\,
    {\cal L}_{\bar\theta\xi}^{(1/2)}(y)\,{\cal J}_M^{(0)}(0) \right\} .
\end{equation}
Even after the decoupling transformation (\ref{blocks}) this operator 
contains arbitrarily complicated soft-collinear exchanges, as illustrated 
in the cartoon in Figure~\ref{fig:SCblob}.

Note that the superficially leading term in the time-ordered product 
cancels \cite{Becher:2003qh}. This can be understood as follows. After 
the decoupling transformation the strong-interaction part of the 
SCET$_{\rm II}$ Lagrangian no longer contains unsuppressed interactions 
between soft-collinear messengers and soft or collinear fields. In order 
to preserve a transparent power counting it is then convenient to define 
hadron states in the effective theory as eigenstates of one of the two 
leading-order Lagrangians ${\cal L}_s$ and ${\cal L}_c$. For instance, we 
define a ``SCET pion'' to be a bound state of only collinear fields, and 
a ``SCET $B$ meson'' to be a bound state of only soft fields. The SCET 
pion state coincides with the true pion, because the collinear Lagrangian 
is equivalent to the QCD Lagrangian \cite{Beneke:2002ph}.\footnote{The 
endpoint region of the pion wave function is not described in terms of a 
Fock component containing a soft-collinear parton, but rather in terms of 
a time-ordered product of the SCET pion state with an insertion of the 
Lagrangian ${\cal L}_{\bar\theta\xi}$. This insertion is non-zero only in 
processes where also soft partons are involved.} 
The SCET $B$ meson coincides with the asymptotic heavy-meson state as 
defined in heavy-quark effective theory. It is important in this context 
that time-ordered products of soft-collinear fields with only collinear 
or only soft fields vanish to all orders as a consequence of analyticity 
\cite{Becher:2003qh}. Hence, soft-collinear modes do not affect the 
spectrum of hadronic eigenstates of the collinear or soft Lagrangians. 
Once the SCET$_{\rm II}$ hadron states are defined in this way, each term 
in the time-ordered product (\ref{3T}) can be factorized into a part 
containing all soft and collinear fields times a vacuum correlation 
function of the soft-collinear fields. These vacuum correlators must be 
invariant under rotations in the transverse plane. It follows that the 
correlator arising from the superficially leading term in the 
time-ordered product vanishes, since
\begin{equation}\label{vacuum}
   \langle\Omega|\,\mbox{T}\,\{ (\bar\sigma\,S_{sc})_i(y_-)\,
   (S_{sc}^\dagger\,W_{sc})_{jk}(0)\,(W_{sc}^\dagger\,\theta)_l(x_+)
   \}\,|\Omega\rangle
\end{equation}
contains a single transverse derivative (see (\ref{sigmasol})).

The leading terms in the time-ordered product in (\ref{3T}) scale like
$\lambda^4$, since ${\cal J}_M^{(0)}\sim\lambda^{5/2}$. They thus 
contribute at the same order in power counting to the universal functions 
$\zeta_M$ as the four-quark operators discussed in 
Section~\ref{sec:matching}. When $O_5^{(M)}$ is added to the list of 
four-quark operators a complete description of the soft-overlap 
contribution is obtained. The Wilson coefficient of this new operator 
follows from the fact that there is no non-trivial matching coefficient 
in (\ref{Jmatch}), that the current operators have the same anomalous 
dimensions in SCET$_{\rm I}$ and SCET$_{\rm II}$, and that the 
SCET$_{\rm II}$ Lagrangian is not renormalized. Hence, to all orders in 
perturbation theory
\begin{equation}\label{D5}
   D_5^{(M)}(\mu_{hc},\mu) = \frac{C_i(\mu)}{C_i(\mu_{hc})}
   \equiv D_5(\mu_{hc},\mu) \,,
\end{equation}
which is in fact a universal function, independent of the labels ``$i$''
and ``$M$''. As before, $\mu_{hc}$ denotes the hard-collinear matching 
scale, at which the transition SCET$_{\rm I}$\,$\to$\,SCET$_{\rm II}$ is
made. The appearance of $C_i(\mu_{hc})$ in the denominator of this 
relation is due to the fact that this coefficient was factored out in the 
definition of $D_k^{(M)}$, see~(\ref{ff}) and (\ref{scet2ff}).

If we define by $\xi_k^{(M)}$ the $B\to M$ hadronic matrix elements of 
the operators $O_k^{(M)}$, then the sum 
$\zeta_M=\sum_k D_k^{(M)}\,\xi_k^{(M)}$ describes the entire soft overlap 
contribution. Each term in this sum gives a contribution of order 
$\lambda^{3/2}$ to the form factors, in accordance with the scaling law
obtained a long time ago in the context of QCD sum rules 
\cite{Chernyak:ag}. We have thus completed the derivation of the new 
factorization formula (\ref{scet2ff}). Whereas the Wilson coefficients 
$D_k^{(M)}$ depend on the renormalization scale $\mu$ as well as on the 
hard-collinear scale $\mu_{hc}\sim\sqrt{E\Lambda}$, the characteristic 
scale of the hadronic matrix elements $\xi_k^{(M)}$ is the QCD scale 
$\Lambda$, not the hard-collinear scale. While this is obvious for the 
matrix elements of the operator $O_5^{(M)}$, it also holds true for the 
remaining matrix elements, for which the sensitivity to long-distance 
physics is signaled by the presence of endpoint singularities. It remains 
to discuss how our results are affected by single and double logarithmic 
corrections arising in higher orders of perturbation theory.

\section{Operator mixing and Sudakov logarithms}
\label{sec:Sudakov}

Because the operators $O_k^{(M)}$ share the same global quantum numbers
they can mix under renormalization. This mixing is governed by a 
$5\times 5$ matrix of anomalous dimension kernels, which are in general
complicated functions of the light-cone variables $u_i$ and $\omega_i$. 
The anomalous dimension matrix governing this mixing has the structure
\begin{equation}\label{gamma}
   \bm{\gamma} = \left( \begin{array}{cccc|c}
    \gamma_{11} & 0 & 0 & \gamma_{14} & 0 \\
    0 & \gamma_{22} & \gamma_{23} & 0 & 0 \\
    0 & \gamma_{32} & \gamma_{33} & 0 & 0 \\
    \gamma_{41} & 0 & 0 & \gamma_{44} & 0 \\
    \hline
    \gamma_{51} & \gamma_{52} & \gamma_{53} & \gamma_{54} & \gamma_{55}
   \end{array} \right) .
\end{equation}
Because the operators $O_{1\dots 4}^{(M)}$ consist of products of soft 
and collinear currents (after decoupling of soft-collinear messengers), 
the entries in the upper left $4\times 4$ sub-matrix can all be written 
as sums of soft and collinear anomalous dimensions for the corresponding 
current operators. We have taken into account that the operators 
$O_2^{(M)}$ and $O_3^{(M)}$ mix under renormalization (this mixing is 
determined by the mixing of twist-3 two-particle and three-particle LCDAs 
for light mesons), as do the operators $O_1^{(M)}$ and $O_4^{(M)}$ (this 
mixing has not yet been worked out but is allowed on general grounds). 
The operator $O_5^{(M)}$ consisting of a triple time-ordered product 
requires the ``local'' operators as counter-terms; however, those 
operators do not mix back into $O_5^{(M)}$.

Because of the structure of the anomalous dimension matrix it follows 
that the coefficient $\gamma_{55}$ is one of the eigenvalues, which
governs the scale dependence of the Wilson coefficient $D_5$ in 
(\ref{D5}). This coefficient coincides with the well-known anomalous 
dimension of the current operators $J_M^{(0)}$ and ${\cal J}_M^{(0)}$
computed in \cite{Bauer:2000yr,Becher:2003kh}. The corresponding 
``operator eigenvector'' can be written as a linear combination (in the 
convolution sense) 
\begin{equation}\label{calO5}
   {\cal O}_5^{(M)} = O_5^{(M)} + \sum_{k=1}^4 d_k^{(M)}
   \otimes O_k^{(M)} \,, \vspace{-0.2cm}
\end{equation}
where the coefficients $d_k^{(M)}$ are independent of the high-energy 
matching scale $\mu_{hc}$. The other four eigenvectors are linear 
combinations of the operators $O_k^{(M)}$ with $k\ne 5$. This observation 
has an important consequence: Since the $B\to M$ matrix elements of the 
operators $O_k^{(M)}$ with $k\ne 5$ are singular, only linear 
combinations of these operators with $O_5^{(M)}$ can contribute to the 
form factors, because only they could be free of such 
singularities.\footnote{Since each eigenvector is multiplied by a Wilson 
coefficient with a different dependence on the high-energy matching 
scale (as determined by its anomalous dimension eigenvalue), no 
accidental cancellations between different eigenvectors can occur.} 
However, the single combination of all five operators that is an 
eigenvector is ${\cal O}_5^{(M)}$. Therefore, we conclude that the 
combination of operators contributing to the universal form factors must, 
up to an overall factor, coincide with this eigenvector, and hence
\begin{equation}
   d_k^{(M)} = \frac{D_k^{(M)}}{D_5}
\end{equation}
to all orders in perturbation theory. This result implies several 
remarkable facts, which could be checked by higher-order perturbative 
calculations. First, since the coefficients $d_k^{(M)}$ must be 
independent of $\mu_{hc}$, each of the coefficients $D_k^{(M)}$ must 
share the same dependence on the high-energy matching scale, and this 
dependence must cancel that of the SCET$_{\rm I}$ Wilson coefficients 
$C_i(E,\mu_{hc})$ in (\ref{ff}). This statement is compatible with the 
observation that the operators $O_k^{(M)}$ arise from the matching of 
time-ordered products of the current $J_M^{(0)}$ with the SCET$_{\rm I}$ 
Lagrangian onto the low-energy theory \cite{Bauer:2002aj}, but it is by 
no means an automatic consequence of this observation. Secondly, it 
follows that there must be intimate relations between the Wilson 
coefficients $D_k^{(M)}$ with $k\ne 5$ and the off-diagonal elements 
$\gamma_{5k}$ of the anomalous dimension matrix in (\ref{gamma}), which 
are connected through the RG equation
\begin{equation}
   \left( \frac{d}{d\ln\mu} + \gamma_{55} \right) d_k
   - d_j\otimes\gamma_{jk} = \gamma_{5k} \,,
\end{equation}
which follows from (\ref{calO5}).

With a slight abuse of notation, let us now denote by $\zeta_M$ the 
$B\to M$ hadronic matrix element of the eigenvector ${\cal O}_5^{(M)}$ in
SCET$_{\rm II}$. Combining (\ref{ff}) and (\ref{D5}), we then find that 
the spin-symmetric universal form-factor term can be rewritten as
\begin{equation}\label{wow}
   C_i(E,\mu_{\rm I})\,\zeta_M(\mu_{\rm I},E) \big|_{\rm SCET_I}
   =  C_i(E,\mu)\,\zeta_M(\mu,E) \big|_{\rm SCET_{II}} \,.
\end{equation}
This relation is not as dull as it seems; rather, it contains the
remarkable message that for the soft overlap contribution to 
heavy-to-light 
form factors the intermediate hard-collinear scale is without {\em any\/} 
physical significance. Switching from SCET$_{\rm I}$ to SCET$_{\rm II}$ 
we merely describe the same physics using a different set of degrees of 
freedom. In other words, there is no use of going through an intermediate 
effective theory. The RG evolution of the soft functions $\zeta_M$ 
remains the same all the way from the high-energy scale $E\sim m_B$ down 
to hadronic scales $\mu\sim\Lambda$. The physics of the soft overlap term 
is thus rather different from the physics of the spin-symmetry breaking 
corrections in the factorization formula (\ref{ff}), for which the 
hard-collinear scale is of physical significance. Any spin-symmetry 
breaking contribution involves at least one hard-collinear gluon 
exchange, and the amplitude factorizes below the scale $\mu_{hc}$.

The result (\ref{wow}) allows us to systematically resum the 
short-distance logarithms arising in the evolution from high energies 
down to hadronic scales. The RG equation obeyed by the Wilson coefficient 
functions is \cite{Bauer:2000yr,Becher:2003kh}
\begin{equation}\label{RGE}
   \frac{d}{d\ln\mu}\,C_i(E,\mu)
   = \left( \Gamma_{\rm cusp}[\alpha_s(\mu)]\,\ln\frac{2E}{\mu}
   + \gamma[\alpha_s(\mu)] \right) C_i(E,\mu) \,,
\end{equation}
where the coefficient of the logarithmic term is determined by the cusp 
anomalous dimension \cite{Korchemsky:wg}. Its solution is
\begin{equation}\label{Cisol}
   C_i(E,\mu) = C_i(E,\mu_h)\,\exp U(\mu_h,\mu,E) \,,
\end{equation}
where $\mu_h\sim 2E$ is the high-energy matching scale for the transition 
from QCD to SCET, at which the values of the Wilson coefficients can be 
reliably computed using fixed-order perturbation theory. The RG evolution 
function can be written as
\begin{equation}\label{Uevol}
   U(\mu_h,\mu,E)
   = \int\limits_{\alpha_s(\mu_h)}^{\alpha_s(\mu)}\!d\alpha\,
    \frac{\Gamma_{\rm cusp}(\alpha)}{\beta(\alpha)}
    \Bigg[ \ln\frac{2E_\gamma}{\mu_h}
    - \int\limits_{\alpha_s(\mu_h)}^\alpha
    \frac{d\alpha'}{\beta(\alpha')} \Bigg]
   + \int\limits_{\alpha_s(\mu_h)}^{\alpha_s(\mu)}\!d\alpha\,
    \frac{\gamma(\alpha)}{\beta(\alpha)} \,,
\end{equation}
where $\beta(\alpha_s)=d\alpha_s/d\ln\mu$. The dependence on the 
high-energy matching scale $\mu_h$ cancels against that of the Wilson 
coefficients $C_i(E,\mu_h)$ in (\ref{Cisol}). Note that after 
exponentiation the evolution function contains an energy and 
scale-dependent factor $\exp U(\mu_h,\mu,E)\propto E^{a(\mu)}$, where
\begin{equation}
   a(\mu) = \int^{\alpha_s(\mu)}\!d\alpha\,
   \frac{\Gamma_{\rm cusp}(\alpha)}{\beta(\alpha)} \,.
\end{equation} 
In order for this scale dependence to be canceled, the low-energy 
hadronic  matrix element must carry an energy dependence of the form 
$(\Lambda/E)^{a(\mu)}$.

Evaluating the hadronic matrix elements at a low scale means that all 
{\em short-distance\/} dependence on the large energy $E$ is extracted 
and resummed in the coefficient functions. However, in the present case 
the matrix elements still contain a long-distance dependence on the large 
scale $E$, which cannot be factorized \cite{Becher:2003kh}. The reason is 
that the large energy enters the effective theory as an external variable 
imprinted by the particular kinematics of soft-to-collinear transitions. 
Because of the large Lorentz boost $\gamma=E/m_M$ connecting the rest 
frames of the $B$ meson and the light meson $M$, the low-energy effective 
theory knows about the large scale $E$ even though hard quantum 
fluctuations have been integrated out. This is similar to applications of 
heavy-quark effective theory to $b\to c$ transitions, where the fields 
depend on the external velocities of the hadrons containing the heavy 
quarks, and $\gamma=v_b\cdot v_c$ is an external parameter that appears 
in the matrix elements of velocity-changing current operators 
\cite{Falk:1990yz,Neubert:1993mb}. In the present case, it is perhaps 
reasonable to assume that the primordial energy dependence of the 
hadronic matrix elements at some low hadronic scale might be moderate, so
that the dominant $E$ dependence is of short-distance nature and can be 
extracted into the Wilson coefficient functions. However, there will 
always be some energy dependence left in the matrix elements; even if we 
assume that it is absent for some value of $\mu$, it will unavoidably be 
reintroduced when we change the scale. As a result, it is impossible to 
determine the asymptotic behavior of the QCD form factors 
$f_i^{B\to M}(q^2)$ using short-distance methods.\footnote{The same 
phenomenon is known to occur in the case of the Sudakov form factor, for 
which the coefficient of the double logarithm is sensitive to infra-red 
physics \cite{Korchemsky:1988hd,Kuhn:1999nn}.}

Let us now proceed to study the numerical importance of short-distance 
Sudakov logarithms. Given the exact results in (\ref{Cisol}) and 
(\ref{Uevol}), it is straightforward to derive approximate expressions 
for the resummed Wilson coefficients at a given order in RG-improved 
perturbation theory by using perturbative expansions of the anomalous 
dimensions and $\beta$-function. Unfortunately, controlling terms of 
$O(\alpha_s)$ in the evolution function $U$ would require knowledge of 
the cusp anomalous dimension to three-loop order (and knowledge of 
$\gamma$ to two-loop order), which at present is lacking. We can, 
however, control the dependence on the recoil energy $E$ to 
$O(\alpha_s)$. Following \cite{Bosch:2003fc}, we define the ratio 
$r=\alpha_s(\mu)/\alpha_s(\mu_h)$ and obtain
\begin{equation}
   e^{U(\mu_h,\mu,E)} = e^{U_0(\mu_h,\mu)}
   \left( \frac{2E}{\mu_h} \right)^{-\frac{\Gamma_0}{2\beta_0} \ln r}
   \left[ 1 - \frac{\alpha_s(\mu_h)}{4\pi}\,\frac{\Gamma_0}{2\beta_0}
   \left( \frac{\Gamma_1}{\Gamma_0} - \frac{\beta_1}{\beta_0} \right)
   (r-1)\,\ln\frac{2E}{\mu_h} \right] ,
\end{equation}
where
\begin{eqnarray}
   U_0(\mu_h,\mu) 
   &=& \frac{\Gamma_0}{4\beta_0^2} \left[
    \frac{4\pi}{\alpha_s(\mu_h)} \left( 1 - \frac{1}{r} - \ln r \right)
    + \frac{\beta_1}{2\beta_0}\,\ln^2 r
    - \left( \frac{\Gamma_1}{\Gamma_0} - \frac{\beta_1}{\beta_0} \right) 
    (r-1-\ln r) \right] \nonumber\\
   &&\mbox{}- \frac{\gamma_0}{2\beta_0}\,\ln r + O(\alpha_s) \,.
\end{eqnarray}
The only piece missing for a complete resummation at 
next-to-next-to-leading order is the $O(\alpha_s)$ contribution to $U_0$, 
which is independent of $E$. The relevant expansion coefficients are 
$\Gamma_0=\frac{16}{3}$, 
$\Gamma_1=\frac{1072}{9}-\frac{16}{3}\pi^2-\frac{160}{27}n_f$, 
$\gamma_0=-\frac{20}{3}$, and $\beta_0=11-\frac23\,n_f$, 
$\beta_1=102-\frac{38}{3}n_f$. We set $n_f=4$ in our numerical work.

\begin{figure}
\begin{center}
\epsfig{file=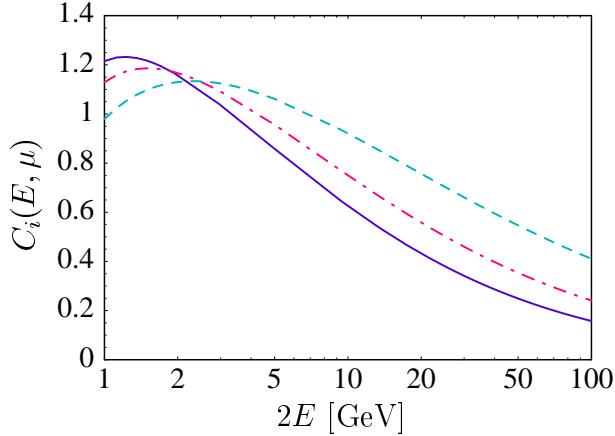,width=8cm} 
\end{center}
\centerline{\parbox{14cm}{\caption{\label{fig:Sudakov}
Energy dependence of the Wilson coefficients $C_i(E,\mu)$ at 
next-to-leading order in RG-improved perturbation theory. The three 
curves correspond to $\alpha_s(\mu)=1$ (solid), $\alpha_s(\mu)=0.75$ 
(dashed-dotted), and $\alpha_s(\mu)=0.5$ (dashed), which are 
representative of typical hadronic scales.}}}
\end{figure}

In Figure~\ref{fig:Sudakov} we show the dependence of the Wilson 
coefficients $C_i(E,\mu)$ on the large energy $E$. We choose $\mu_h=2E$ 
for the high-energy matching scale and use the tree-level initial
conditions $C_i(E,\mu_h)=1$, which is consistent at next-to-leading 
order. We fix $\mu$ at a low hadronic scale in order to maximize the 
effect of Sudakov logarithms. The maximum recoil energy in $B\to\pi$ 
transitions is such that $2E_{\rm max}\simeq 5.3$\,GeV. Obviously, for 
such values the perturbative resummation effects are very moderate. In 
the energy range $1\,\mbox{GeV}<2E<m_B$ the Wilson coefficients differ 
from unity by no more than about 20\%. The extrapolation to larger energy 
values shows that Sudakov suppression remains a moderate effect even for 
very large recoil energy.

\section{Factorization of spin-symmetry breaking effects}
\label{sec:fact}

Using the close connection between messenger exchange and endpoint 
singularities discussed in Section~\ref{sec:toy}, the formalism of 
soft-collinear fields can be used to demonstrate the convergence of 
convolution integrals in QCD factorization theorems. If messenger 
exchange is unsuppressed, the convolution integrals diverge at the 
endpoints, spoiling factorization. By the same reasoning, convolution 
integrals are finite if soft-collinear messenger contributions are absent 
at leading power. 

Let us apply this method to show, to all orders in perturbation theory,
that the convolution integrals entering the spin-symmetry breaking term
in the factorization formula (\ref{ff}) are free of endpoint 
singularities. As mentioned in the Introduction, this is an essential
ingredient still missing from the proof of this formula. With our 
technology the proof is rather simple and consists of only the following 
two steps:

1.\;
Soft-collinear messenger fields can be decoupled from the four-quark
operators mediating spin-symmetry breaking effects, which are of the type 
shown in (\ref{Q1Q2}). The reason is that in the color singlet-singlet 
case the operators are invariant under the field redefinition 
(\ref{blocks}) \cite{Becher:2003kh}. It follows that their matrix 
elements factorize into separate matrix elements of soft and collinear 
currents, which can be written in terms of the leading-order LCDAs of the 
external mesons. If the messengers did not decouple, they would introduce 
unsuppressed interactions between the soft and collinear parts of the 
four-quark operators, thereby spoiling factorization.

2.\;
Time-ordered products containing interactions of soft-collinear 
messengers with soft or collinear fields do not contribute to the 
spin-symmetry violating term in (\ref{ff}). This follows from SCET power 
counting. The insertions of the Lagrangians ${\cal L}_{\bar q\theta}$ and 
${\cal L}_{\bar\theta\xi}$ in (\ref{3T}) cost a factor $\lambda^{3/2}$,
meaning that they can only come together with a leading-order current
operator of the type $\bar\X\dots\H\sim\lambda^{5/2}$. (Note that there 
are no $O(\lambda^{1/2})$ corrections to the current that could make the 
vacuum correlator (\ref{vacuum}) involving 
${\cal L}_{\bar q\theta}^{(1/2)}$ and 
${\cal L}_{\bar\theta\xi}^{(1/2)}$ non-zero \cite{Becher:2003kh}.) 
However, such a current will always be of the form (\ref{Jmatch}) because 
of the projection properties of the heavy-quark and collinear-quark 
spinors. It thus respects the spin-symmetry relations. 

These two observations imply that at leading order in $\Lambda/E$ 
soft-collinear exchanges do not contribute to the spin-symmetry breaking 
term in the factorization formula, and hence the corresponding 
convolution integrals are convergent.

\section{Summary and conclusions}
\label{sec:concl}

We have presented a complete scale separation for the soft universal form
factors $\zeta_M(E)$, which parameterize the leading spin-symmetric 
contributions to heavy-to-light form factors at large recoil energy. 
These functions are related to hadronic matrix elements of operators in 
soft-collinear effective theory (SCET$_{\rm II}$). When 
evaluated using the standard (four-dimensional) projections onto meson
light-cone distribution amplitudes, these matrix elements times their 
respective Wilson coefficient functions lead to singular convolution 
integrals over light-cone momentum fractions. We have shown that this is 
an indication that exceptional momentum contributions corresponding to 
highly asymmetric Fock states of the external mesons cannot be neglected 
at leading power. In the effective theory, these contributions are 
parameterized in terms of matrix elements of operators involving 
soft-collinear messenger modes, which communicate between the soft and 
collinear sectors of the theory. Using a toy model, we have demonstrated 
that messenger exchange precisely accounts for the endpoint 
configurations in the full theory.

The sum of all effective-theory matrix elements is free of spurious
endpoint singularities, but it cannot be factorized into convolutions of 
hard-scattering kernels with light-cone distribution amplitudes. We have 
argued that the sum of all operators contributing at leading order in 
power counting is an eigenvector of the anomalous dimension matrix 
governing operator mixing in the effective theory. The corresponding 
eigenvalue coincides with the anomalous dimension of the leading-order 
SCET$_{\rm II}$ current operator containing a heavy quark and a collinear 
quark. Using this result, we have performed a complete resummation of 
short-distance Sudakov logarithms arising in the evolution from the hard 
scale $\mu\sim 2E\sim m_B$ down to a low hadronic scale $\mu\sim\Lambda$. 
For the physical value of the $B$-meson mass, the resummed Sudakov factor 
leads to no suppression at all. Even for much larger values 
$m_B\approx 50$--100\,GeV, the suppression would only be a factor 2--3. 
This is insufficient to suppress the soft overlap contribution. The main 
conclusion of our work is therefore that the soft overlap contribution to 
heavy-to-light form factors exists, and that it does not receive a 
significant perturbative suppression. 

A perhaps surprising finding of our analysis is that the intermediate 
hard-collinear scale $\sqrt{E\Lambda}$ is irrelevant to the physics of 
the soft overlap contribution. This is, at first sight, a puzzling 
result, because the hard-collinear scale seems to set the natural scale 
for the gluon virtualities in Feynman graphs contributing to the form 
factors, and more generically it is (in many cases) the characteristic 
scale for the interactions of soft and collinear degrees of freedom. 
Also, at the hard-collinear scale one usually switches between different 
effective field theories (called SCET$_{\rm I}$ and SCET$_{\rm II}$) and 
hence describes physics using different sets of degrees of freedom. 
Nevertheless, we have shown that the evolution of the soft functions 
$\zeta_M(E)$ remains the same all the way from the high-energy scale 
$\mu\sim m_B$ down to hadronic scales $\mu\sim\Lambda$. The physics of 
the soft overlap term is thus rather different from the physics of the 
spin-symmetry breaking corrections in the factorization formula 
(\ref{ff}), for which the hard-collinear scale is of physical 
significance.

Another unusual (although in retrospect not surprising) result of our 
study is the observation that the hadronic matrix elements parameterizing 
the soft overlap contribution contain a long-distance dependence on the 
large scale $E$, which cannot be factorized using short-distance methods. 
The large energy enters the effective theory as an external variable 
imprinted by the particular kinematics of soft-to-collinear transitions, 
which are characterized by a large Lorentz boost connecting the rest 
frames of the $B$ meson and the light meson $M$. The best that can be 
achieved is to extract all {\em short-distance\/} dependence on the 
recoil energy by evaluating the matrix elements at a low scale. While it 
is reasonable to assume that the primordial energy dependence of the 
hadronic matrix elements at some low hadronic scale might be moderate, it 
is impossible to determine the precise asymptotic behavior of the QCD 
form factors $f_i^{B\to M}$ using short-distance methods.

These results have important phenomenological implications. First, the 
fact that the soft overlap contribution is unsuppressed implies that the
spin-symmetric contributions to the form factors are parametrically 
larger than spin-symmetry violating corrections, which are suppressed by 
a factor of $\alpha_s(\sqrt{E\Lambda})$. This lends credibility to the 
idea of an approximate spin symmetry realized in the large-energy limit 
of QCD. Secondly, our finding that the soft overlap is not significantly 
suppressed by a short-distance Sudakov factor casts doubt on one of the 
key assumptions underlying the pQCD approach to exclusive $B$ decays 
\cite{Keum:2000ph}. It supports the power counting scheme employed in the 
QCD factorization approach, where form factors are treated as 
non-perturbative hadronic input parameters 
\cite{Beneke:1999br}. Finally, it should be obvious from our discussion 
that the physics underlying heavy-to-light form factors at large recoil 
is subtle and far more complicated than the ``wave-function overlap'' 
picture usually associated with simpler form factors such as those 
relevant to $K\to\pi$ or $B\to D$ transitions. We feel that the physics 
associated with the transformation of the soft degrees of freedom in the 
$B$ meson into the energetic partons of the light final-state meson is 
not properly accounted for in existing model calculations of 
heavy-to-light form factors using quark models or QCD sum rules. A more 
detailed investigation of the numerical implications of our findings is 
left for future work.

\vspace{0.5cm}\noindent
{\em Acknowledgments:\/}
We are grateful to Thomas Becher, Thorsten Feldmann, Richard Hill and 
Peter Lepage for many clarifying discussions. This research was supported 
by the National Science Foundation under Grant PHY-0098631.

\end{document}


\begin{center}
\Large\bf 
Erratum:\\ 
Factorization and the soft overlap contribution to heavy-to-light form 
factors
\end{center}

\begin{center}
{\sc Bj\"orn O.~Lange$^{(a)}$ and Matthias Neubert$^{(b)}$}\\
\vspace{0.4cm}

$^{(a)}${\sl Center for Theoretical Physics,\\
Massachusetts Institute of Technology, 
Cambridge, MA 02139, U.S.A.}\\[3mm]
$^{(b)}${\sl Institute for High-Energy Phenomenology, Newman Laboratory\\
for Elementary-Particle Physics, Cornell University,
Ithaca, NY 14853, U.S.A.}
\end{center}

\begin{abstract}
\noindent 
We correct an argument about the $\mbox{SCET}_{\rm I}\to\mbox{SCET}_{\rm II}$ 
matching of the soft-overlap contribution to heavy-to-light form factors at 
large recoil.
\end{abstract}
\vspace{0.4cm}

\noindent
In Section~6 entitled ``Operator mixing and Sudakov logarithms'' of our work
\cite{Lange:2003pk}, we stated incorrectly that only linear combinations of 
the operators $O_k^{(M)}$ that include $O_5^{(M)}$ can contribute to the 
$B\to M$ form factors. The argument that led to this conclusion, presented 
after eq.~(47), is erroneous. Linear combinations of the operators $O_k^{(M)}$ 
with $k=1,\dots,4$ may also contribute, provided that the corresponding Wilson 
coefficients are free of endpoint singularities. As a result, eq.~(48) in 
\cite{Lange:2003pk} is incorrect, and eq.~(50) must be generalized to read
\begin{eqnarray}
   C_i(E,\mu_I)\,\zeta_M(\mu_I,E) \big|_{\rm SCET_I}
   &=& C_i(E,\mu)\,\zeta_M(\mu,E) \big|_{\rm SCET_{II}} \\
   &&\hspace{-1.38cm}\mbox{}+ \sum\limits_{k=1}^4 \left[
    C_i(E,\mu_{hc})\,D_k^{(M)}(E,\mu_{hc},\mu) - C_i(E,\mu)\,d^{(M)}_k(E,\mu)
    \right] \otimes\,\xi_k^{(M)}(\mu) \,, \nonumber
\end{eqnarray}
where $\xi_k^{(M)}(\mu)$ denote the factorizable $B\to M$ matrix elements of 
the operators $O_k^{(M)}$ in eqs.~(21) and (22), which can be expressed in 
terms of twist-2 and twist-3, two- and three-particle light-cone distribution 
amplitudes of the $B$ meson and the light meson $M$. The main conclusion of 
our paper, that there exists an unsuppressed ``soft overlap'' contribution to 
the form factors (the first term on the right-hand side in the above equation) 
for which the hard-collinear scale $\mu_{hc}\sim\sqrt{E\Lambda}$ is 
irrelevant, remains unchanged. The coefficient functions in the second line 
of eq.~(50), on the other hand, in general do contain logarithms of the 
hard-collinear scale. An explicit calculation of these coefficients could help 
to shed light on the question about the numerical size of the soft overlap 
piece.

\bigskip\noindent
{\em Acknowledgements:\/}
We are grateful to M.~Beneke and I.~Stewart for useful discussions.